\documentclass[12pt]{article}

\usepackage[utf8]{inputenc}

\usepackage{amsmath,amsthm,amssymb,amscd}
\usepackage{a4wide}

\usepackage{url}

\usepackage{graphicx}
\usepackage[bf,small]{caption}
\usepackage{enumitem}

\usepackage{sectsty}
\sectionfont{\bf\large}
\subsectionfont{\bf\normalsize}

\renewcommand{\Re}{\operatorname{Re}}
\renewcommand{\Im}{\operatorname{Im}}
\newcommand{\Tr}{\operatorname{Tr}}

\usepackage{color}

\definecolor{grey}{gray}{0.4}

\definecolor{mygr}{rgb}{0,0.4,0.4}

\usepackage{def}

\begin{document}

\noindent
{\large\bf Particle creation and annihilation at interior boundaries:\\[.5ex]
One-dimensional models}\\[2ex]
{\bf Stefan Keppeler}${}^a$ and {\bf Martin Sieber}${}^b$\\[2ex]
${}^a$Fachbereich Mathematik, 
Universität Tübingen, 
Auf der Morgenstelle 10, 
72076 Tübingen, 
Germany, 
\url{stefan.keppeler@uni-tuebingen.de}\\
${}^b$School of Mathematics, University of Bristol, University Walk, Bristol, BS8\;1TW, UK,
\url{m.sieber@bristol.ac.uk}\\[2ex]

\parbox{.8\textwidth}{
  {\bf Abstract.} We describe creation and annihilation of particles
  at external sources in one spatial dimension in terms of
  interior-boundary conditions (IBCs). We derive explicit solutions
  for spectra, (generalised) eigenfunctions, as well as Green
  functions, spectral determinants, and integrated spectral
  densities. Moreover, we introduce a quantum graph version of
  IBC-Hamiltonians.}

\section{Introduction}
\label{sec:intro}

Quantum field theories are plagued by infinities. Among the most
serious infinities are ultraviolet divergences. Usually they are taken
care of by renormalisation, often within perturbation theory. However,
in many cases it is not clear whether the renormalised theory exists
at all as a well-defined theory in its own right.

Recently, Teufel and Tumulka proposed a novel formulation of quantum
field theory~\cite{TeuTum15a}, see also \cite{TeuTum15b}, where
particle creation and annihilation is modelled in terms of conditions
coupling Fock space sectors with different numbers of particles. Since
such a condition typically relates the $n$-particle wave function to
the value of the $(n{+}1)$-particle wave function (or it's derivative)
at a specific point, it is called interior-boundary condition (IBC) in
Ref.~\cite{TeuTum15a}. In simple models the IBC formulation is
automatically ultraviolet finite.

In particular, Teufel and Tumulka study models in three spatial
dimensions in which non-relativistic scalar particles can be created
and annihilated at fixed external sources. These models, when
described with the methods of conventional quantum field theory, are
known to be renormalisable, even non-perturbatively~\cite{Der03}. The
IBC-version turns out to be equivalent to the renormalised theory up
to a trivial shift in energy. This correspondence, which is explored in
\cite{TeuTum15a}, is made mathematically rigorous by Lampart, Schmidt,
Teufel and Tumulka \cite{LamSchTeuTum??}, see also \cite{Sch14}, who
show that the relevant IBC-Hamiltonian is essentially self-adjoint and
bounded from below. To this end the authors of
\cite{TeuTum15a,LamSchTeuTum??} have to define the domain of the
single-particle Hamiltonian such that it contains certain singular
functions which are not in the Soboloev space $H^2(\R^3)$. In
particular, they allow simple poles at the positions of the sources.

Teufel and Tumulka also give the IBC-formulation of a model with
dynamical sources, i.e.\ a model in which one kind of particles can be
created at the positions of particles of a different kind. In a
realistic quantum field theory one should think, e.g., of photons
being created at the positions of electrons. It appears that a
rigorous non-perturbative analysis of this latter model can be carried
out along similar lines as for the model with fixed sources
\cite{LamSchTeuTum??b}.

Conditions similar to IBCs have been studied earlier under the name
``zero radius potentials with internal structure'', see
e.g. \cite{PavShu90} and references therein. In this context they are
used to account for rearrangements within a scatterer in diffractive
processes. The alternative interpretation of a particle interacting
with the vacuum has also been given \cite{Yaf92}. This latter
interpretation is more in line with the way in which we mainly want to
interpret these conditions in the present work.

In this article we study IBCs in one spatial dimension as model
systems for which many questions can be answered by explicit
calculations. As opposed to the three-dimensional case, we do not have
to allow single-particle wave functions with poles but only with kinks. 
Moreover, one-dimensional IBCs can also be studied on quantum graphs,
multiply connected one-dimensional systems, which have become
paradigmatic for studies of quantum chaos over the last one and a half
decades, see e.g.\ \cite{GnuSmi06}. The present work thus also
introduces model systems for studying quantum chaos in the context of
many particle quantum mechanics and quantum field theory.

The article is organised as follows. In Sec.~\ref{sec:ibc-hamiltonian}
we motivate the one-dimensional IBC-Hamiltonian as an analogue to the
recently introduced IBCs in three dimensions. We introduce versions on
both, the full bosonic Fock space and on truncated Fock space with a
maximum number of particles. Being interested in the minimal model
exhibiting particle creation and annihilation due to interior-boundary
conditions, 0-1-particle systems get particular attention in the
following sections. In Sec.~\ref{sec:one_source_on_R} we construct a
complete orthonormal set of (generalised) eigenfunctions for the
IBC-Hamiltonian with one source and determine the corresponding
(retarded) Green function. Section~\ref{sec:several_sources} is
devoted to the case of two and more sources. We analyse how the ground
state energy depends on the distance of the sources recovering a
one-dimensional Coulomb potential for small distances. In
Sec.~\ref{sec:source_in_box} we discuss the spectrum for one source in
a finite box and Dirichlet boundary conditions. Quantum graphs with
IBCs in the vertices are introduced in Sec.~\ref{sec:graphs}.

\section{The IBC-Hamiltonian}
\label{sec:ibc-hamiltonian}

Before introducing the IBC-Hamiltonian we briefly recall the
definition of (bosonic and fermionic) Fock space and particle
creation and annihilation operators.

Starting from a one-particle Hilbert space $\mathcal{H}$, say
$\mathcal{H}=L^2(\R)$, Fock space is constructed by taking direct sums
of tensor products of $\mathcal{H}$,
\begin{equation}
  \mathcal{F} = \bigoplus_{n=0}^\infty \mathcal{H}^{\otimes n} \, .
\end{equation}
We represent a vector $\phi \in \mathcal{F}$ as a sequence $\phi =
(\phi^0,\phi^1,\phi^2,\hdots)$ with
\begin{equation}
  \phi^0 \in \mathcal{H}^{\otimes 0} \cong \C \, , \quad
  \phi^1 \in \mathcal{H} \, , \quad
  \phi^2 \in \mathcal{H} \otimes \mathcal H \, , \ \hdots
\end{equation}
We symmetrise in order to obtain bosonic Fock space
$\mathcal{F}^S=P^S(\mathcal{F}) \subset \mathcal{F}$, where the
symmetrisation operator acts on $\phi \in \mathcal{F}$ as
\begin{equation}
  (P^S \phi)^n(x_1,\hdots,x_n) = \frac{1}{n!} \sum_{\sigma \in S_n} 
    \phi^n(x_{\sigma(1)},\hdots,x_{\sigma(n)}) \, .
\end{equation}
Here $S_n$ is the symmetric group of degree $n$.  

Later we are also interested in truncated Fock space for describing
situations with at most $N$ particles,
\begin{equation}
  \mathcal{F}_N 
  = \bigoplus_{n=0}^N \mathcal{H}^{\otimes n} 
  \subset \mathcal{F} \, ,
\end{equation}
and truncated bosonic Fock space $\mathcal{F}^S_N = P^S(\mathcal{F}_N)$. 

Fermionic Fock $\mathcal{F}^A$ space is introduced analogously, with
symmetrisation $P^S$ replaced by anti-symmetrisation $P^A$ acting on
$\phi \in \mathcal{F}$ as 
\begin{equation}
  (P^A \phi)^n(x_1,\hdots,x_n) = \frac{1}{n!} \sum_{\sigma \in S_n} 
  \mathrm{sign}(\sigma) \, \phi^n(x_{\sigma(1)},\hdots,x_{\sigma(n)}) \, .
\end{equation}

Scalar product and norm on $\mathcal{F}$ are induced from the scalar
product on the one-particle Hilbert space $\mathcal{H}$, in particular
for $\mathcal{H}=L^2(\R)$ and $\phi,\psi \in \mathcal{F}$ we have
\begin{equation}
  \langle \phi, \psi \rangle
  = \overline{\phi^0} \, \psi^0 
    + \sum_{n=1}^\infty \int_{-\infty}^\infty 
      \overline{\phi^n(x_1,\hdots,x_n)} \, \psi^n(x_1,\hdots,x_n) 
      \, \ud x_1 \hdots \ud x_n \, .
\end{equation}
On $\mathcal{F}$ one can define operators $a(f)$ which annihilate a
particle with normalised wave function $f$, by
\begin{equation}
\label{eq:a(f)def}
\begin{split}
  \big(a(f)\phi\big)^n(x_1,\hdots,x_n) 
  &= \sqrt{n+1} \int_{-\infty}^\infty \overline{f(x)} \, 
     \phi^{n+1}(x_1,\hdots,x_n,x) \, \ud x 
     \quad \forall\, n \geq 0 \, .
\end{split}
\end{equation}
Note that $a(f)$ leaves the bosonic and fermionic subspaces
$\mathcal{F}^S$ and $\mathcal{F}^A$ invariant. The adjoint of $a(f)$
on $\mathcal{F}^S$, the bosonic creation operator $a^\dag(f)$, acts as 
\begin{equation}
\label{eq:a+(f)def}
\begin{split}
  \big(a^\dag(f)\phi\big)^n(x_1,\hdots,x_n) 
  &= \frac{1}{\sqrt{n}} \sum_{j=1}^n f(x_j) \, 
     \phi^{n-1}(x_1,\hdots,x_{j-1},x_{j+1},\hdots,x_n) 
     \quad \forall\, n \geq 1 \\
  \text{and} \qquad \big(a^\dag(f)\phi\big)^0 &= 0 \, .
\end{split}
\end{equation}
On $\mathcal{F}^A$, i.e.\ for fermions, one has to include an
additional factor $(-1)^{j+1}$ inside the sum.

Now we would like to describe bosons which can be created and
annihilated at an external source located at position $y$ and which
otherwise move freely. Free propagation is described by
$-\Delta^\mathcal{F}$ (in units where $\hbar=2m=1$) which is defined
as
\begin{equation}
  \big(\Delta^\mathcal{F}\phi\big)^n = \sum_{j=1}^n \Delta_j \phi^n
\end{equation}
where $\Delta_j$ denotes the second derivative with respect to the
$j^\text{th}$ argument. Thus, our tentative Hamiltonian reads
\begin{equation}
\label{eq:tentative_H}
  H = -\Delta^\mathcal{F} + \overline{c} \, a(\delta_y) 
                          + c \,  a^\dag(\delta_y) \, ,  
\end{equation}
where $\delta_y(x)=\delta(x-y)$ denotes the Dirac delta function and
$c\in\C$ is a coupling constant. However, since $\delta_y$ is not a
smooth function but a distribution, the creation operator
$a^\dag(\delta_y)$ cannot even be densely defined on $\mathcal{F}^S$,
cf.\ e.g.\ \cite[Sec.~X.7]{ReeSim75-II}, i.e.\ as it stands
Eq.~\eqref{eq:tentative_H} does not make sense. But we can try to give
meaning to Eq.~\eqref{eq:tentative_H} in the same way as
$\delta$-potentials are treated in textbook quantum mechanics. To this
end we write out the eigenvalue equation $H\phi=E\phi$ in the
$n$-particle sector ($n\geq 1$),
\begin{equation}
\label{eq:eigenvalue_n-particle-sector}
\begin{split}
  -\sum_{j=1}^n \big(\Delta_j \phi^n\big)(x_1,\hdots,x_n) 
  &+ \overline{c} \sqrt{n+1} \, \phi^{n+1}(x_1,\hdots,x_n,y) \\
  &\hspace{-10ex}
   + \frac{c}{\sqrt{n}} \sum_{j=1}^n \delta(x_j-y) \, 
     \phi^{n-1}(x_1,\hdots,x_{j-1},x_{j+1},\hdots,x_n) 
  = E \, \phi^n(x_1,\dots,x_n) \, , 
\end{split}
\end{equation}
integrate in one variable, say $x_n$, from $y-\varepsilon$ to
$y+\varepsilon$, and take the limit $\varepsilon\to0+$. We obtain
\begin{equation}
\label{eq:first_IBC}
  -\Big[\big(\partial_n\phi^n\big)(x_1,\hdots
                                   ,x_n)\Big]_{x_n=y-}^{x_n=y+}
  + \frac{c}{\sqrt{n}} \, \phi^{n-1}(x_1,\hdots,x_{n-1}) = 0 \, , 
\end{equation}
where $\partial_j$ denotes the derivative with respect to the $j$th
argument. We have thus found a condition coupling neighbouring sectors
in Fock space, which replaces the ill-defined creation operator in
Eq.~\eqref{eq:tentative_H}. Following Teufel and Tumulka
\cite{TeuTum15a,TeuTum15b} we will refer to Eq.~\eqref{eq:first_IBC}
as an interior-boundary condition (IBC). Rewriting
Eq.~\eqref{eq:first_IBC}, for $c\neq0$ our model now reads
\begin{equation}
\begin{split}
\label{eq:IBC-model_Fock}
  &H = -\Delta^\mathcal{F} + \overline{c} \, a(\delta_y) \\
  \text{ with IBC } \ & 
  \phi^n(x_1,\hdots,x_n) = \frac{\sqrt{n+1}}{c} 
  \Big[\big(\partial_{n+1}\phi^{n+1}\big)(x_1,\hdots,x_{n+1})
  \Big]_{x_{n+1}=y-}^{x_{n+1}=y+} 
  \, .
\end{split}
\end{equation}
This is the one-dimensional analogue of the condition which is called
Dirichlet-IBC in Ref.~\cite{TeuTum15a}, as we demonstrate in
Appendix~\ref{app:3Dibcs}. In \cite{LamSchTeuTum??} it is shown that
the IBC-Hamiltonian is essentially self-adjoint when defined on a
suitable domain. Models with several point sources can be written down
in the same way, by adding additional annihilation operators to the
Hamiltonian, supplemented by the corresponding IBCs, see also
Sec.~\ref{sec:several_sources}. Inspection of the IBC in
Eq.~\eqref{eq:IBC-model_Fock} reveals that $\phi^n$ uniquely determines
$\phi^\nu$ $\forall\, \nu < n$. Moreover, the IBC ensures that
$\phi^n$ inherits the symmetry of $\phi^{n+1}$, i.e.\ although derived
for bosons, only a factor $(-1)^n$ inside the square brackets is
required in order to define the corresponding IBC-Hamiltonian for
fermions.

The IBC-Hamiltonian~\eqref{eq:IBC-model_Fock} is, up to a translation
in energy, unitarily equivalent to the free Hamiltonian
$-\Delta^\mathcal{F}$, a result which we will discuss elsewhere. (The
analogous statement for the 3D IBC-Hamiltonian is shown in
\cite{LamSchTeuTum??,Sch14}.) In
Secs.\ \ref{sec:one_source_on_R}--\ref{sec:source_in_box} we instead
focus on the truncated IBC-Hamiltonian on $\mathcal{F}^S_1$, i.e.\ we
do not allow creation of more than one particle. Correspondingly,
$\phi=(\phi^0,\phi^1)$ and the IBC-Hamiltonian reads
\begin{equation}
\label{eq:1D-IBC-Hamiltonian}
  (H\phi)^1 = -{\phi^1}'' \, , \quad 
  (H\phi)^0 = \overline{c} \phi^1(0) \quad \text{with IBC} \quad
  \phi^0 = \frac{1}{c} \Big[{\phi^1}'(x)\Big]_{x=0-}^{x=0+} \, , 
\end{equation}
where we have, without loss of generality, also specialised to $y=0$.

Teufel and Tumulka emphasise that for their choice of IBC (in three
dimensions) probability is not conserved within Fock space sectors
with fixed numbers of particles, but that it is conserved on full Fock
space. In particular, the IBC enables probability flow between Fock
space sectors with different numbers of particles. The same is true
for the one-dimensional IBC models of Eqs.~\eqref{eq:IBC-model_Fock}
and \eqref{eq:1D-IBC-Hamiltonian} which can be seen as follows.
Consider, e.g., the time dependent Schrödinger equation
$\ui\dot{\phi}=H\phi$ with Hamiltonian and IBC
\eqref{eq:1D-IBC-Hamiltonian}, i.e.\ sectorwise we have
\begin{align}
\label{eq:Schrodinger_1sec}
  \ui\dot{\phi^1}(x,t) &= -{\phi^1}''(x,t) \, , \quad 
  x \neq 0 \quad \text {and}
  \\
\label{eq:Schrodinger_0sec}
  \ui\dot{\phi^0}(t) &= \overline{c} \phi^1(0,t) \, .
\end{align}
From Eq.~\eqref{eq:Schrodinger_1sec} one can derive the continuity
equation $\dot{\rho^1}(x,t)+{j^1}'(x,t)=0$, $x\neq0$, in the
one-particle sector, with probability density $\rho^1(x,t) =
|\phi^1(x,t)|^2$ and current $j^1(x,t) =
2\Im\big(\overline{\phi^1(x,t)}{\phi^1}'(x,t)\big)$ as usual. However,
at $x=0$, the position of the source, the current is generally
discontinuous. According to the IBC the flow into the origin is
\begin{equation}\label{flux1->0}
 \left[-j^1(x,t) \right]_{x=0-}^{x=0+} 
 = -2 \Im\big(c\overline{\phi^1(0,t)}\phi^0(t)\big) \, .
\end{equation}
This is compensated by the continuity equation in the zero-particle
sector, 
\begin{equation}\label{eq:continuity-0sec}
  \frac{\ud}{\ud t}|\phi^0(t)|^2
  = -2\Im\big(c\overline{\phi^1(0,t)}\phi^0(t)\big) \, , 
\end{equation}
which is readily derived from Eq.~\eqref{eq:Schrodinger_0sec}.  Notice
that Eqs.~\eqref{flux1->0} and \eqref{eq:continuity-0sec} imply that
for stationary states, i.e.\ for solutions of the time-independent
Schrödinger equation $H\phi=E\phi$, there can be no net flux between
the zero-particle and the one-particle sector.

\section{Spectrum and (generalised) eigenfunctions for one source}
\label{sec:one_source_on_R}

We consider the eigenvalue equation $H\phi=E\phi$ for the
one-dimensional IBC-Hamilto\-ni\-an~\eqref{eq:1D-IBC-Hamiltonian} with
$x\in\R$. As $\phi^1$ has to solve $-{\phi^1}''(x)=E\phi^1(x)$ for
$x\neq0$ we distinguish the cases of negative energy, $E<0$, and
positive energy, $E>0$. 
 
\vspace{1ex}\noindent {\bf Ground state ($E<0$).} 
For negative energy we make the ansatz $\phi^1(x) = A
\ue^{-\kappa|x|}$ with $\kappa>0$, which solves the eigenvalue
equation in the one-particle sector with $E=-\kappa^2$. Combining the
eigenvalue equation in the zero-particle sector, $E
\phi^0=\overline{c}A$, with the IBC, $\phi^0=-2A\kappa/c$, yields the
condition $2\kappa^3=|c|^2$. Together with the normalisation condition
\begin{equation}
\left\|\left(\phi_\mathrm{g}^0,\phi_\mathrm{g}^1 \right)\right\|^2
= |A|^2 \frac{4\kappa^2}{|c|^2} + |A|^2 \int_{-\infty}^\infty \ue^{-2\kappa|x|} \, \ud x
= |A|^2 \left( \frac{2}{\kappa} + \frac{1}{\kappa} \right) = 1 \, ,
\end{equation}
we find $A = \sqrt{\kappa/3}$ and obtain the normalised ground state
\begin{equation} \label{phig}
\begin{split} 
  \phi_\mathrm{g} = (\phi_\mathrm{g}^0, \phi_\mathrm{g}^1) \, , \quad 
  \phi_\mathrm{g}^0 = -\sqrt{\tfrac{2}{3}} \, \tfrac{|c|}{c} \, , \quad 
  \phi_\mathrm{g}^1(x) = \sqrt{\tfrac{\kappa}{3}} \, \ue^{-\kappa|x|} 
  \, , \quad
  \kappa = \sqrt[3]{\tfrac{|c|^2}{2}}\, , 
\end{split}
\end{equation}
with energy $E_\mathrm{g}=-\kappa^2$. Note that the state has a weight
of $2/3$ in the zero-particle sector and a weight of $1/3$ in the
one-particle sector
\begin{equation} \label{phignorm}
\overline{\phi_\mathrm{g}^0} \phi_\mathrm{g}^0 = \frac{2}{3} 
\, , \qquad
\int_{-\infty}^\infty \overline{\phi_\mathrm{g}^1(x)} \, \phi_\mathrm{g}^1(x) 
\, \ud x = \frac{1}{3} \, , 
\end{equation}
with sum $\|\phi_\mathrm{g}\|^2 = \langle \phi_\mathrm{g} ,
\phi_\mathrm{g} \rangle = 1$. In limit $c\to0$ $\phi_\mathrm{g}$
becomes proportional to the free vacuum $\phi_\mathrm{vac} =
(\phi_\mathrm{vac}^0,\phi_\mathrm{vac}^1) = (1,0)$, but the
normalisation is lost. In appendix \ref{app:recover_ground_state} we
show how $\phi_\mathrm{vac}$ can be recovered from $\phi_g$ in the
limit of vanishing coupling by first introducing a zero point energy.

In the following it is often convenient to express $c$ in terms of
$\kappa$ and write $c=\sqrt{2\kappa^3}\,\ue^{\ui\varphi_c}$ where
$\varphi_c$ is the phase of $c$.

\vspace{1ex}\noindent 
{\bf Scattering states (generalised eigenfunctions for $E>0$).} 
In the one-particle sector we make the ansatz 
\begin{equation} \label{phik1}
  \phi_k^1(x) = \frac{1}{\sqrt{2 \pi}} \left( 
    \ue^{\ui kx} + b_k \ue^{\ui |k||x|} \right) \, , \quad 
  k \in \R \setminus\{0\} \, , 
\end{equation}
of a (flux normalised) plane wave plus a one-dimensional ``spherical''
wave. For $x\neq0$ we have $-\Delta \phi_k^1 = k^2 \phi_k^1$,
i.e.\ the energy of the tentative solution is $E_k=k^2$. From the IBC
follows
\begin{equation} \label{phik0}
  \phi_k^0 = \frac{2\ui |k|}{\sqrt{2 \pi} \, c} \, b_k \, , 
\end{equation} 
and together with $(H\phi)^0=E\phi^0$ we obtain 
\begin{equation} \label{bk}
\frac{\overline{c}}{\sqrt{2 \pi}} (1+b_k) = E \frac{2\ui |k|}{\sqrt{2 \pi} \, c} \, b_k 
\qquad \Leftrightarrow \qquad 
b_k = \frac{|c|^2}{2\ui|k|^3-|c|^2} = \frac{- \ui \kappa^3}{|k|^3 + \ui \kappa^3} \, . 
\end{equation}
The scattering states \eqref{phik1}, being solutions of the
time-independent Schrödinger equation, conserve the probability flux
within the one-particle sector, cf.\ the remark following
Eq.~\eqref{eq:continuity-0sec}. In the language of scattering theory, this
can also be seen by decomposing $\phi_k^1$ into an incoming wave, a
reflected wave with reflection amplitude $r_k=b_k$, and a transmitted
wave with transmission amplitude $t_k=1+b_k$. Then flux conservation
within the one-particle sector means that reflection and transmission
coefficients add to unity, $|r_k|^2+|t_k|^2=1$, which is equivalent to
$\Re b_k = -|b_k|^2$, a condition fulfilled by \eqref{bk}.

Notice that $\lim\limits_{c\to0}b_k=0$ as well as $\lim\limits_{c\to0}
\phi_k^0 = 0$, i.e.\ for vanishing coupling the solution $\phi_k =
(\phi_k^0, \phi_k^1)$ goes over to an ordinary plane wave supported
only in the one-particle sector.

The scattering states \eqref{phik1} with coefficients $b_k$ from
\eqref{bk} are very similar to those for a system with delta potential
in single-particle quantum mechanics: A particle subject to a
potential $\lambda \delta(x)$, but otherwise free, has scattering
states of the form \eqref{phik1} with $b_k$ replaced by the reflection
amplitude $\lambda/(2 \ui |k| - \lambda)$ \cite{AGHH88}. Hence the
scattering solutions \eqref{phik1} formally look like those for a
delta potential with (energy-dependent) parameter $\lambda = |c|^2/E$.

Ground state and scattering states form an orthonormal set
satisfying
\begin{equation} \label{on}
\langle \phi_g , \phi_g \rangle = 1 \, , \qquad \langle \phi_g , \phi_k \rangle = 0 \, , \qquad
\langle \phi_{k'} , \phi_k \rangle = \delta(k-k') \, , 
\end{equation}
for all $k, k' \in \mathbb{R} \setminus \{0\}$. The first relation was
obtained above and the other two are derived in
Appendix~\ref{sec:orthocomplete}.

\vspace{1ex}\noindent {\bf Completeness}. 
Having established orthonormality we now move on to show that
$(\phi_\mathrm{g}^0,\phi_\mathrm{g}^1)$ and $(\phi_k^0,\phi_k^1)$ form
a complete set. Completeness means that any $\phi=(\phi^0,\phi^1) \in
\mathcal{F}^S_1$ can be expanded as follows,
\begin{equation}
\label{eq:expansion_into_basis}
  \phi = a_\mathrm{g} \phi_\mathrm{g} + \int_{-\infty}^\infty a_k
  \phi_k \, \ud k \, .
\end{equation}
Due to orthonormality the expansion coefficients can be calculated by
taking scalar products,
\begin{equation}
  a_\mathrm{g} = \langle\phi_\mathrm{g},\phi\rangle  \, , \qquad 
  a_k = \langle \phi_k,\phi\rangle \, .
\end{equation} 
Inserting the coefficients into the expansion,
Eq.~\eqref{eq:expansion_into_basis} becomes, sectorwise,
\begin{equation}
\begin{split}
  \phi^0 
  &= \left( |\phi_\mathrm{g}^0|^2 
            + \int_{-\infty}^\infty |\phi_k^0|^2 \, \ud k \right) \phi^0 
   + \int_{-\infty}^\infty \left( 
     \overline{\phi_\mathrm{g}^1(y)} \, \phi_\mathrm{g}^0 
     + \int_{-\infty}^\infty \overline{\phi_k^1(y)} \, \phi_k^0 \, \ud k \right)
     \phi^1(y) \,  \ud y \, , \\
  \phi^1(x)
  &= \left( \overline{\phi_\mathrm{g}^0} \, \phi_\mathrm{g}^1(x)  
     + \int_{-\infty}^\infty \overline{\phi_k^0} \, \phi_k^1(x) \, \ud k \right)
     \phi^0 \\ & \quad\
     + \int_{-\infty}^\infty \left( 
     \overline{\phi_\mathrm{g}^1(y)} \, \phi_\mathrm{g}^1(x) 
     + \int_{-\infty}^\infty \overline{\phi_k^1(y)} \, \phi_k^1(x) \, \ud k
     \right) \phi^1(y) \, \ud y \, .
\end{split}
\end{equation} 
These equations hold for every $\phi$ if and only if 
\begin{enumerate}
\item[(i)]
$\displaystyle |\phi_\mathrm{g}^0|^2 + \int_{-\infty}^\infty |\phi_k^0|^2 \, \ud k = 1$,
\item[(ii)] 
$\displaystyle \overline{\phi_\mathrm{g}^0} \, \phi_\mathrm{g}^1(x) 
 + \int_{-\infty}^\infty \overline{\phi_k^0} \, \phi_k^1(x) \, \ud k = 0$
\quad and
\item[(iii)]
$\displaystyle \overline{\phi_\mathrm{g}^1(y)} \, \phi_\mathrm{g}^1(x) 
 + \int_{-\infty}^\infty \overline{\phi_k^1(y)} \, \phi_k^1(x) \, \ud k 
 = \delta(x-y)$.
\end{enumerate}
These three relations are shown to hold for the solutions
$\phi_\mathrm{g}$ and $\phi_k$ in Appendix~\ref{sec:orthocomplete}.

\vspace{1ex}\noindent {\bf Green function.} 
The resolvent $(E-H)^{-1}$ of the IBC-Hamiltonian on $\mathcal{F}^S_1$
has the spectral decomposition
\begin{equation} \label{resolventspectral}
\frac{1}{E-H} = \frac{\langle \phi_\mathrm{g} , \cdot \rangle \phi_\mathrm{g}}{E - E_\mathrm{g}}
+ \int_{-\infty}^\infty \frac{\langle \phi_k , \cdot \rangle \phi_k}{E - E_k} \, \ud k \, .
\end{equation}
If we express the resolvent in the position representation then we
obtain the Green function.  Letting $(E-H)$ act on the Green function
one obtains from \eqref{eq:1D-IBC-Hamiltonian} and
\eqref{resolventspectral} the following equations in the different
sectors
\begin{alignat}{2} \label{green_equations}
& \left( E + \frac{\ud^2}{\ud x^2} \right) \, G^{11}(x,y,E) = \delta(x-y) \, , \qquad && E \, G^{01}(y,E) - \overline{c} G^{11}(0,y,E) = 0 \, , \notag \\
& \left( E + \frac{\ud^2}{\ud x^2} \right) \, G^{10}(x,E) = 0 \, , \qquad  && E\, G^{00}(E) - \overline{c} G^{10}(0,E) = 1 \, ,
\end{alignat}
for $x \neq 0$. They are subject to the IBCs
\begin{equation} \label{green_ibcs}
\begin{split}
G^{01}(y,E) & = \frac{1}{c} \left[ \frac{\ud}{\ud x} G^{11}(x,y,E) \right]_{x=0-}^{x=0+} \, , \quad
G^{00}(E) = \frac{1}{\overline{c}} \left[ \frac{\ud}{\ud y} G^{01}(y,E) \right]_{y=0-}^{y=0+} \, , \\
G^{10}(x,E) & = \frac{1}{\overline{c}} \left[ \frac{\ud}{\ud y} G^{11}(x,y,E) \right]_{y=0-}^{y=0+} \, , \quad
G^{00}(E) = \frac{1}{c} \left[ \frac{\ud}{\ud x} G^{10}(x,E) \right]_{x=0-}^{x=0+} \, .
\end{split}
\end{equation}

The solutions to \eqref{green_equations} and \eqref{green_ibcs} can be
obtained directly from the scattering states \eqref{phik1}.  Let
$k=\sqrt{E}>0$. Then the (retarded) Green function is given in the
$11$-sector by
\begin{equation} \label{green_wronskian}
G^{11}(x,y,E) = \frac{\phi_{-k}^1(x_<) \, \phi_k^1(x_>)}{W(y)} \, ,
\end{equation}
where $x_>$ ($x_<$) is the larger (smaller) of $x$ and $y$, and $W(y)
= \phi_{-k}^1(y) \, {\phi_{k}^1}'(y) - \phi_{k}^1(y) \,
{\phi_{-k}^1}'(y)$ is the Wronskian. Formulas of the form
\eqref{green_wronskian} are used in Sturm-Liouville theory, see
e.g.\ \cite[Sec.~7.2]{MorFes53}. The term $\phi_{-k}^1(x_<)$ has the correct
outgoing behaviour as $x \rightarrow -\infty$, whereas
$\phi_{k}^1(x_>)$ is outgoing for $x \rightarrow \infty$.  With
\eqref{phik1} we obtain $W = 2 \ui k (1 + b_k)$, and from
\eqref{green_wronskian}
\begin{equation} \label{G11}
G^{11}(x,y,E) = \frac{1}{2 \ui k} \, \ue^{\ui k |x-y|} + \frac{b_k}{2 \ui k} \, \ue^{\ui k |x|} \, \ue^{\ui k |y|} \, .
\end{equation}
Again this has a similar form as for a delta potential
$\lambda\delta(x)$, see \cite{AGHH88}. As before, the difference is
that $|c|^2/E$ replaces $\lambda$. The Green function in the other
sectors follows from \eqref{green_ibcs},
\begin{equation} \label{G01G10G00}
G^{01}(y,E) = \frac{b_k}{c} \, \ue^{\ui k |y|} \, , \qquad
G^{10}(x,E) = \frac{b_k}{\overline{c}} \, \ue^{\ui k |x|} \, , \qquad
G^{00}(E) = \frac{2 \ui k b_k}{|c|^2} \, .
\end{equation}
It can easily be checked that these solutions satisfy the relations
\eqref{green_equations}. For negative energies one has to set $k=\ui
\sqrt{-E}$.

In order to develop an intuitive interpretation of the Green function,
we define a diffraction coefficient $\mathcal{D} = 2 \ui k b_k$ which
allows us to rewrite Eq.~\eqref{G11} as
\begin{equation} \label{green_gtd}
G^{11}(x,y,E) = G_0(x,y,E) + G_0(x,0,E) \, \mathcal{D} \, G_0(0,y,E) \, .
\end{equation}
Here $G_0(x,y,E) = \exp(\ui k |x{-}y|) / 2 \ui k$ is the free
single-particle Green function in one dimension. In the semiclassical
limit $k \rightarrow \infty$ the Green function can be interpreted in
terms of particle trajectories, and in the case at hand the
semiclassical approximation coincides with the exact Green
function. In \eqref{green_gtd} there is a contribution from a direct
path from $y$ to $x$, and a second path from $y$ to the origin where
it is diffracted due to the interaction with the vacuum and then
continues to $x$. This agrees with the general form expected according
to Keller's geometrical theory of diffraction
\cite{Kel62,VWR94}. Since for the IBC-Hamiltonian the only
interactions possible at the origin are particle creation and
annihilation, we would like to interpret $\mathcal{D}$ as the
amplitude for annihilation and subsequent re-creation of a particle.
Consequently, $G^{11}$ displays flux conservation in the one-particle
sector. For Green functions this is expressed by the optical theorem
which here takes the form $\Im \mathcal{D} = - |{\cal D}|^2/2k$, cf.\
\cite{BLS00}.

Before we can interpret the Green function in all sectors in terms of
particle creation and annihilation we need to introduce one more
notion. In the same way as we refer to $G_0(x,y,E)$ as the amplitude
for a particle going from $y$ to $x$ along a straight line, let us
call $G^{00}(E)$ the amplitude for staying in the vacuum. Now we can
once more rewrite the Green function,
\begin{equation}
\begin{split}
  G^{11}(x,y,E) &= G_0(x,y,E) 
    + G_0(x,0,E) \, c \, G^{00}(E) \, \overline{c} \, G_0(0,y,E)\\[.5ex]
G^{01}(y,E) &= G^{00}(E) \, \overline{c} \, G_0(0,y,E) \, , \\[.5ex]
G^{10}(x,E) &= G_0(x,0,E) \, c \, G^{00}(E) \, .
\end{split}\end{equation}
Recall that according to the IBC-Hamiltonian particle creation is
associated with a factor $c$, whereas particle annihilation is
associated with a factor $\overline{c}$. This is made most obvious in
the naive Hamiltonian~\eqref{eq:tentative_H}. Now all components of
the Green function can be interpreted in the same way. $G^{01}(y,E)$
is the amplitude for a particle going from $y$ to the origin where it
is annihilated, and then the system stays in the vacuum. The reverse
process in described by $G^{10}(x,E)$, with the system starting in the
vacuum, then a particle is created and moves to $x$. Finally, the
diffractive contribution to $G^{11}(x,y,E)$ can be interpreted as the
amplitude for a particle going from $y$ to the origin where it is
annihilated, intermediately leaving the system in the vacuum, and then
the particle is re-created and moves to $x$.

\vspace{1ex}\noindent {\bf Time evolution.} 
Stationary solutions at constant energy $E$ cannot have a net
probability flux between the zero-particle and the one-particle
sector. The situation is different for the time evolution of general
states.  Consider a state $\phi(t)$ satisfying the initial condition
$\phi(0)=\phi_0\in\mathcal{F}^S_1$. For times $t>0$ it can be
expressed in the two sectors in terms of the initial state
$\phi_0=(\phi_0^0,\phi_0^1)$ and the kernel $K$ of the time evolution
operator,
\begin{align}
\phi^1(x,t) & = \int_{-\infty}^\infty K^{11}(x,y,t) \, \phi_0^1(y) \, \ud y + K^{10}(x,t) \, \phi^0_0 \, , \notag \\
\phi^0(t) & = \int_{-\infty}^\infty K^{01}(y,t) \, \phi_0^1(y) \, \ud y + K^{00}(t) \, \phi^0_0 \, .
\end{align}
The time evolution kernel for the IBC-Hamiltonian
\eqref{eq:1D-IBC-Hamiltonian} is derived in appendix
\ref{sec:propagator} and given explicitly in Eqs.~\eqref{k11final} and
\eqref{k10k01k00}. We consider here only one example. Assume that the
initial state is the free vacuum $\phi_0 = \phi_{\mathrm{vac}} =
(\phi_{\mathrm{vac}}^0, \phi_{\mathrm{vac}}^1) = (1,0)$. For $t>0$ it
evolves to
\begin{equation} 
\phi^1(x,t) = K^{10}(x,t) \, , \qquad \phi^0(t) = K^{00}(t) \, .
\end{equation}
At initial time $t=0$ the state has weight 1 in the zero-particle
sector. Then an interesting question is how the weight changes as $t
\rightarrow \infty$. This follows from the asymptotics of $K^{00}(t)$
as $t \rightarrow \infty$ which is evaluated in \eqref{k00asympt},
\begin{equation}
\phi^0(t) = K^{00}(t) \sim \frac{2}{3} \, \ue^{\ui \kappa^2 t} 
  + \mathcal{O}(t^{-3/2}) \quad \text{as} \quad t \rightarrow \infty \, .
\end{equation}
Hence the weight in the zero-particle sector decreases from $1$ to $4/9$
as $t \rightarrow \infty$.

\section{Several sources}
\label{sec:several_sources}

In this section, we first consider the case of two sources on the real
line. We are particularly interested in the ground state energy and
its dependence on the distance of the sources. Placing the sources at
$y_1=0$ and $y_2=R$, the model reads
\begin{equation}
\begin{split}
  & (H\phi)^1 = -{\phi^1}'' \, , \qquad 
  (H\phi)^0 = \overline{c}_1 \phi^1(0) + \overline{c}_2 \phi^1(R)\\[1ex] 
  \quad \text{with IBCs} \qquad 
  &\phi^0 = \frac{1}{c_1} \Big[{\phi^1}'(x)\Big]_{x=0-}^{x=0+} \, , \qquad 
  \phi^0 = \frac{1}{c_2} \Big[{\phi^1}'(x)\Big]_{x=R-}^{x=R+} \, ,
\end{split}
\end{equation}
and with coupling constants $c_1$ and $c_2$. Thinking of the sources
as particles of a different kind pinned at a distance $R$, the
constants $c_1$ and $c_2$ represent the charges, with which these
pinned particles couple to the dynamical particles of our
model. Hence, the dependence of the ground state energy on the
distance of the sources should be interpreted as the potential between
the charges generated by the exchange of particles in
$\mathcal{F}^S_1$.

\begin{figure}[t]
\centerline{\includegraphics[width=1.13\textwidth]{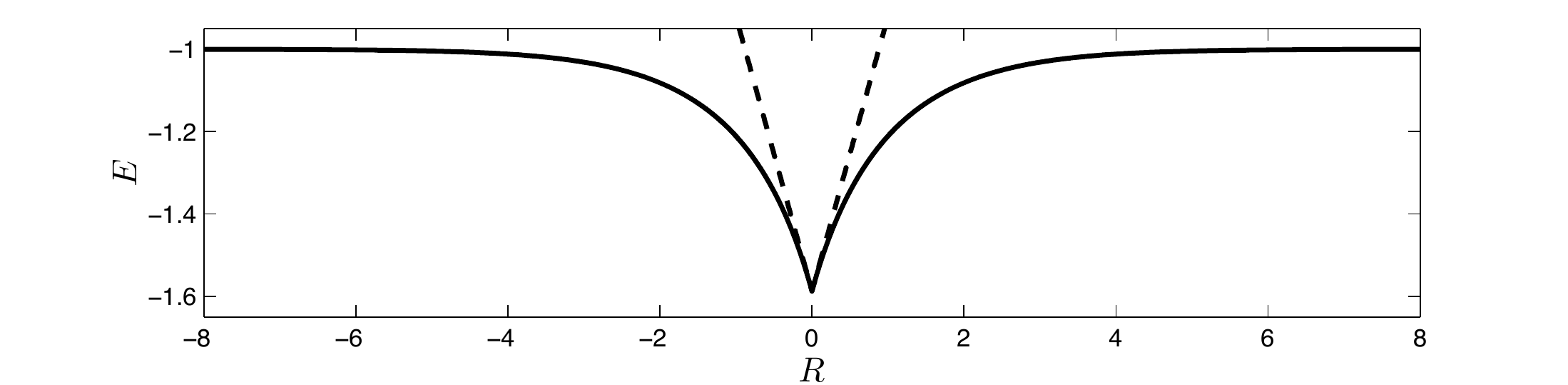}}
\caption{Ground state energy for two sources with coupling constants
  $c_1=c_2=1$ and separation $R$. The dashed lines show the linear
  approximation \eqref{eq:ground_state_energy_small_dist}.}
\label{fig:ground_state-energy_2_sources}
\end{figure}

\vspace{1ex}\noindent {\bf Ground state.} 
As we are interested in the ground state energy we search for
solutions to $H\phi=E\phi$ with negative energy $E<0$. Away from the
sources, $\phi^1$ has to satisfy $-{\phi^1}''=E\phi^1$, and the most
general ansatz for a continuous normalisable solution with negative
energy and discontinuous derivatives at $x=0$ and $x=R$ is
\begin{equation}
  \phi^1(x) = a \, \ue^{-\kappa|x|} + b \, \ue^{-\kappa|x-R|} \, , 
  \quad \kappa > 0 \, , \quad E=-\kappa^2 \, . 
\end{equation}
The IBCs allow to express the coefficients $a$ and $b$ in terms of the
zero-particle sector wave function $\phi^0$, 
\begin{equation}
  a=-\frac{c_1}{2\kappa} \, \phi^0 \, , \qquad 
  b=-\frac{c_2}{2\kappa} \, \phi^0 \, .
\end{equation}
Then the eigenvalue equation in the zero-particle sector reads
\begin{equation}
\label{eq:ground_state-energy_2_sources}
  - \left( \frac{|c_1|^2+|c_2|^2}{2\kappa} 
           + \frac{\overline{c_1} c_2 + \overline{c_2} c_1}{2\kappa} \, 
             \ue^{-\kappa|R|} \right) \phi^0 
  = E \phi^0 \, .
\end{equation}
This defines the ground state energy $E(R)$ as a function of the
distance of the two sources. Substituting $\kappa=\sqrt{-E}$ we first
observe that
\begin{equation}
  E(0) = - \left( \frac{|c_1+c_2|^2}{2} \right)^{2/3}
  \qquad \text{and} \qquad
  \lim_{|R|\to\infty} E(R) = - \left( \frac{|c_1|^2+|c_2|^2}{2} \right)^{2/3}
  \, .
\end{equation}
For small $|R|$ we can expand the exponential in
Eq.~\eqref{eq:ground_state-energy_2_sources} and find the linear
approximation
\begin{equation}
\label{eq:ground_state_energy_small_dist}
  E(R) \sim - \left( \frac{|c_1+c_2|^2}{2} \right)^{2/3} 
            + \frac{\overline{c_1} c_2 + \overline{c_2} c_1}{3} \, |R|
  \, , \quad |R| \to 0 \, ,
\end{equation}
i.e.\ for small distances the ground state energy behaves like a
one-dimensional Coulomb potential.  With real charges $c_1$ and $c_2$
the potential is attractive if the charges have the same sign, and
repulsive for opposite signs, which is to be expected for a scalar
field. In Fig.~\ref{fig:ground_state-energy_2_sources} we display
$E(R)$ for $c_1=c_2=1$ along with the approximation for small $|R|$
(dashed line).

\vspace{1ex}\noindent {\bf Scattering states.} 
For the scattering states we make the ansatz
\begin{equation}
  \phi_k^1(x) = \frac{1}{\sqrt{2 \pi}} \left( 
  \ue^{\ui kx} + b_k \ue^{\ui |k||x|} + \tilde{b}_k \ue^{\ui |k||x-R|} \right) 
  \, , \quad k \in \R \setminus\{0\} \, .
\end{equation}
These functions satisfy $-{\phi_k^1}'' = k^2 \phi_k^1$ for $x \notin
\{0,R\}$ and have energy $E_k=k^2$. From the IBCs follows
\begin{equation} \label{phi0bb}
\phi_k^0 = \frac{2 \ui |k| \, b_k}{\sqrt{2 \pi} \, c_1} \qquad \text{and} \qquad \phi_k^0 = \frac{2 \ui |k| \, \tilde{b}_k}{\sqrt{2 \pi} \, c_2} \, ,
\end{equation}
and from $(H\phi)^0=E\phi^0$ we obtain 
\begin{equation}
E \phi_k^0 = \frac{\overline{c_1}}{\sqrt{2 \pi}} \left( 1 + b_k + \tilde{b}_k \ue^{\ui |k| |R|} \right) 
+ \frac{\overline{c_2}}{\sqrt{2 \pi}} \left( \ue^{\ui k R}  + b_k  \ue^{\ui |k| |R|} + \tilde{b}_k \right) \, .
\end{equation}
Inserting \eqref{phi0bb} we find
\begin{equation}
\phi_k^0 = \frac{2 \ui |k| \, (\overline{c_1} + \overline{c_2} \ue^{\ui k R})}{\sqrt{2 \pi} (2 \ui |k|^3 - |c_1|^2 - |c_2|^2
- (c_1 \overline{c_2} + \overline{c_1} c_2) \ue^{\ui |k| |R|})} \, ,
\end{equation}
and $b_k$ and $\tilde{b}_k$ follow from \eqref{phi0bb}. In the case of
just one source we were able to map (generalised) eigenfunctions to
solutions of a single-particle Schrödinger equation with delta
scatterer, see Sec.~\ref{sec:one_source_on_R}. For two sources it is
no longer possible to map the amplitudes $b_k$ and $\tilde{b}_k$ to
the corresponding amplitudes for two delta scatterers in
single-particle quantum mechanics. The reason for this is that the two
sources do not act as independent scatterers but are connected by the
vacuum.

\vspace{1ex}\noindent {\bf More than  two sources.} 
The results of this section can easily be generalised to several
sources. Then one has
\begin{equation}
(H\phi)^1 = -{\phi^1}'' \, , \quad 
(H\phi)^0 = \sum_{i=1}^n \overline{c}_i \phi^1(x_i) \, , \quad
\phi^0 = \frac{1}{c_i} \Big[{\phi^1}'(x)\Big]_{x=x_i-}^{x=x_i+} \, , \quad 
i=1, \ldots , n \, ,
\end{equation}
for $n$ sources with coupling constants $c_i$ at positions $x_i$. The
ground state has the form
\begin{equation}
  \phi^1_g(x) = \sum_{i=1}^n a_i \ue^{- \kappa |x-x_i|} \, , \quad \kappa > 0 \, , \quad E = - \kappa^2 \, ,
\end{equation}
from the IBCs one obtains
\begin{equation}
  a_i = - \frac{c_i}{2 \kappa} \, \phi^0_g \, , \quad i=1,\ldots,n \, ,
\end{equation}
and the energy $E = -\kappa^2$ follows from the eigenvalue equation in
the zero-particle sector
\begin{equation}
2 \kappa^3 = \sum_{i,j=1}^n \overline{c}_i c_j \ue^{- \kappa \, |x_i - x_j|} \, .
\end{equation}
The scattering states have the form
\begin{equation}
  \phi_k^1(x) = \frac{1}{\sqrt{2 \pi}} \left( 
    \ue^{\ui kx} + \sum_{i=1}^n b_i \ue^{\ui |k||x-x_i|} \right) 
  \, , \quad k \in \R \setminus\{0\} \, , \quad E = k^2 \, ,
\end{equation}
from the IBCs one obtains
\begin{equation}
  b_i = \sqrt{2 \pi} \frac{c_i}{2 \ui |k|} \, \phi^0_k \, , \quad i=1,\ldots,n \, ,
\end{equation}
and from the eigenvalue equation in the zero-particle sector follows
\begin{equation}
\phi^0_k = \frac{2 \ui |k| \sum_{i=1}^n \overline{c}_i \ue^{\ui k x_i}}{\sqrt{2 \pi} \left(2 \ui |k|^3 - \sum_{i,j=1}^n \overline{c}_i c_j \ue^{\ui |k| \, |x_i - x_j|} \right)} \, .
\end{equation}

\section{Particle in a box with one source}
\label{sec:source_in_box}

An example of a system with a discrete spectrum is a particle in a box
of length $l$ with one source.  We place the source at $x=0$ and
require Dirichlet boundary conditions at the end points $x=-l_1<0$ and
$x=l_2>0$ where $l_1 + l_2=l$.

\vspace{1ex}\noindent {\bf Negative energies.}  
The general ansatz for an eigenfunction with energy $E=-\kappa^2$,
$\kappa>0$, which satisfies Dirichlet conditions at $-l_1$ and $l_2$
is
\begin{equation} \label{boxnegansatz}
  \phi^1(x) = \left\{ \begin{matrix} 
  A \sinh\big(\kappa(x+l_1)\big) \, , & x<0 \, , \\
  B \sinh\big(\kappa(x-l_2)\big) \, , & x>0 \, . \end{matrix} \right.
\end{equation}
Continuity at $x=0$ requires
\begin{equation} \label{boxconti}
  A \sinh(\kappa l_1) = - B \sinh(\kappa l_2) \, .
\end{equation}
The IBC \eqref{eq:1D-IBC-Hamiltonian} determines the zero-particle sector
wave function,
\begin{equation}
  \phi^0 = \frac{\kappa}{c}\big(B\cosh(\kappa l_2)-A\cosh(\kappa l_1)\big) \, , 
\end{equation}
and together with $(H\phi)^0=\overline{c} A \sinh(\kappa l_1)$ the
eigenvalue equation demands
\begin{equation}
\label{eq:IBC+eigenvalue}
  |c|^2 A \sinh(\kappa l_1) 
  = -\kappa^3 \big( B \cosh(\kappa l_2) - A \cosh(\kappa l_1) \big) \, ,
\end{equation}
which, using \eqref{boxconti}, can be expressed as 
\begin{equation} \label{boxground}
\kappa^3 = |c|^2 \,  \frac{\sinh(\kappa l_1) \, \sinh(\kappa l_2)}{\sinh(\kappa l)}  \, .
\end{equation}
This equation has exactly one positive solution $\kappa$ for any $l_1,
\, l_2 >0$. This can be seen by considering the left- and right-hand
sides of \eqref{boxground} as functions of $\kappa \ge 0$.  Both
functions start at zero. The left-hand side has a slope that is
monotonously increasing from zero to infinity. The function on the
right-hand side has a slope that is positive and monotonously
decreasing, as we show below. Hence the two functions intersect
exactly once.

To prove the statement about the derivative of the right-hand side
consider
\begin{equation}
\frac{\ud}{\ud \kappa} \frac{\sinh(\kappa l_1) \, \sinh(\kappa l_2)}{\sinh(\kappa l)}
=\frac{l_1 \sinh^2(\kappa l_2) + l_2 \sinh^2(\kappa l_1)}{\sinh^2 (\kappa l)} \, .
\end{equation}
This is positive for $\kappa \ge 0$. It is also monotonously
decreasing because $\sinh(\kappa l_2)/\sinh(\kappa l)$ and
$\sinh(\kappa l_1)/\sinh(\kappa l)$ are monotonously decreasing. This
follows, for example, from
\begin{equation}
\frac{\ud}{\ud \kappa} \frac{\sinh(\kappa l_1)}{\sinh(\kappa l)}
= \frac{l_1 \sinh(\kappa l_2) - l_2 \sinh(\kappa l_1) \cosh(\kappa l)}{\sinh^2(\kappa l)}
< \frac{l_1 \sinh(\kappa l_2) - \kappa l_1 l_2 \cosh(\kappa l_2)}{\sinh^2(\kappa l)} < 0 \, ,
\end{equation}
where the last inequality holds since $\sinh(z) - z \cosh(z) < 0
\ \forall\, z>0$.

\begin{figure}[t]
\centerline{\includegraphics[width=.95\textwidth]{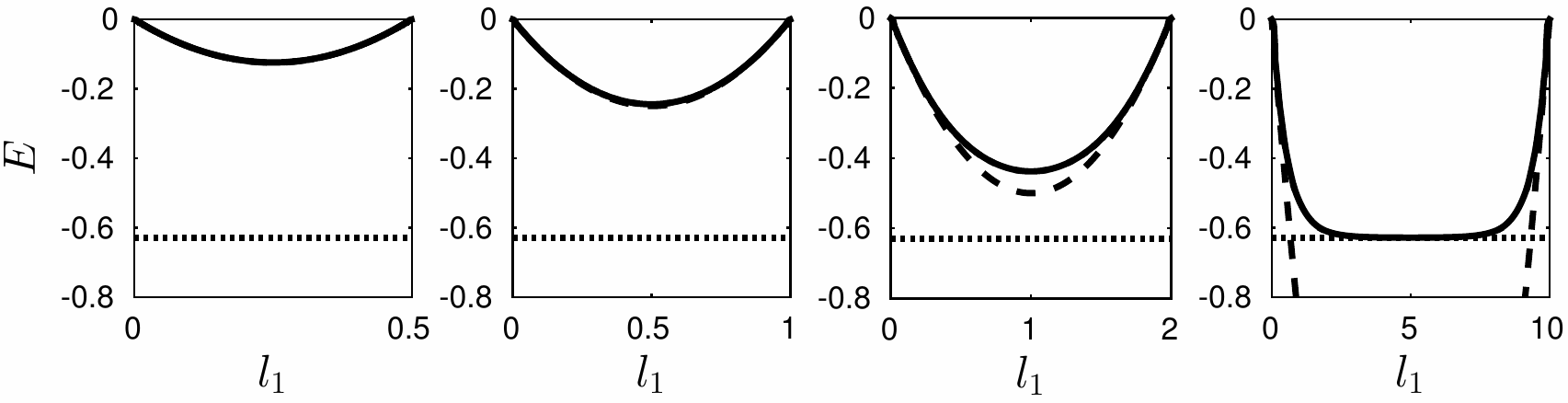}}
\caption{Ground state energy for coupling $|c|=1$ determined by
  Eq.~\eqref{boxground} as a function of the position $l_1$ of the
  source for box lengths $l=\frac{1}{2},1,2,10$ (left to right) along
  with the approximation for small boxes (dashed line, almost
  indistinguishable from the solid line in the first two panels) and
  the limiting value $-2^{-2/3}$ for large boxes (dotted line).}
\label{fig:ground_state_energy_box}
\end{figure}

In the limit of small boxes, $l \rightarrow 0$ with $l_1/l$ and
$l_2/l$ constant, we find from \eqref{boxground} that $\kappa \sim
\sqrt{|c|^2 l_1 l_2/l}$, i.\,e.\ the energy approaches zero from
below. For large boxes, $l \rightarrow \infty$ with $l_1/l$ and
$l_2/l$ constant, we obtain $\kappa \rightarrow \kappa_0 =
\sqrt[3]{|c|^2/2}$. This limiting value corresponds to the ground
state of the IBC-Hamiltonian in Sec.~\ref{sec:one_source_on_R}. In
Fig.~\ref{fig:ground_state_energy_box} we show the ground state energy
as a function of $l_1$, the relative position of the source inside the
box ($0 \leq l_1 \leq l$), for different box sizes and fixed
coupling. We observe an effective repulsion of the source from the
boundaries, generated by emission, reflection and re-absorption of
particles.

\vspace{1ex}\noindent {\bf Zero energy.} 
The general ansatz for an eigenfunction with $E=0$ that satisfies the
Dirichlet boundary conditions is
\begin{equation}
  \phi^1(x) = \left\{ \begin{matrix} 
  A (x+l_1) \, , & x<0 \\
  B (x-l_2) \, , & x>0 \end{matrix} \right. \, . 
\end{equation}
The eigenvalue equation in the zero-particle sector then requires
$A=B=0$, and hence $E=0$ is not an eigenvalue of the Hamiltonian.

\vspace{1ex}\noindent {\bf Positive energies.} 
For eigenfunction with energy $E=k^2$, $k>0$, the ansatz
\eqref{boxnegansatz} is replaced by
\begin{equation} \label{boxposansatz}
  \phi^1(x) = \left\{ \begin{matrix} 
  A \sin\big(k(x+l_1)\big) \, , & x<0 \, , \\
  B \sin\big(k(x-l_2)\big) \, , & x>0 \, . \end{matrix} \right.
\end{equation}
A similar calculation as before then leads to the eigenvalue condition 
\begin{equation} \label{boxhigher}
- k^3  =  |c|^2 \,  \frac{\sin(k l_1) \, \sin(k l_2)}{\sin(k l)} \, .
\end{equation}
In the following we derive spectral determinant and trace formula for
the particle in a box with IBC.  We start by first considering a
particle in a box without IBCs.

\vspace{1ex}\noindent {\bf Green function for box without source.} 
The Green function is obtained from the general formula
\eqref{green_wronskian} by choosing for the left and right functions
$\sin(k (x_< + l_1))$ and $\sin(k(x_> - l_2))$.  These are solutions
of the Schr\"odinger equation that satisfy the left and right
Dirichlet boundary condition, respectively.  With the Wronskian $W(y)
= k \sin(k l)$ we obtain
\begin{equation} \label{green_box}
G_\mathrm{b}(x,y,E) = \frac{\sin \big( k(x_< + l_1) \big) \, \sin \big( k(x_> - l_2) \big)}{k \, \sin(k l)} \, .
\end{equation}
From the poles one reads off the eigenvalue condition $\sin(kl) = 0$,
$k \neq 0$. Note that \eqref{green_box} can also be obtained by
applying mirror images to satisfies the boundary conditions
\begin{equation} \label{green_composite}
G_\mathrm{b}(x,y,E) = \sum_{n=-\infty}^\infty G_0(x,y+2nl,E) - \sum_{n=-\infty}^\infty G_0(x,-y-2l_1+2nl,E) \, ,
\end{equation}
where $G_0$ is the free Green function (after \eqref{green_gtd}). As
discussed earlier, $G_0$ coincides with its semiclassical
approximation in terms of a trajectory from initial to final
point. The form \eqref{green_composite} can then be interpreted as sum
over all trajectories in the box from $y$ to $x$ (unfolded onto the
real line by the mirror principle). This agrees with the general
semiclassical form of Green functions, see e.g.\ \cite{Gut90,Sto99}.

The trace of the Green function is 
\begin{equation}
\Tr G_\mathrm{b}(E) 
= \int_{-l_1}^{l_2} G_\mathrm{b}(x,x,E) \, \ud x 
= \frac{l \cot(k l)}{2 k} - \frac{1}{2 k^2} \, ,
\end{equation}
from which we obtain the spectral determinant
\begin{equation} \label{specdetb}
\Delta_\mathrm{b}(E) 
= \prod_{n=1}^\infty \left( 1 - \frac{E}{E_n^b} \right) 
= \lim_{\varepsilon \rightarrow 0} \exp \left( \int_0^E \Tr G_\mathrm{b}(E + \ui \varepsilon) \, \ud E \right) 
= \frac{\sin(k l)}{k l} \, .
\end{equation}
This is an entire function of $E=k^2$ with zeros at the energy levels
$E_n^\mathrm{b} = (n \pi / l)^2$, $n\in\N$.  The spectral staircase $N(E)$
can be obtained from the spectral determinant by
\begin{equation} \label{staircase_def}
N(E) = \sum_{n=1}^\infty \Theta(E - E_n) = - \frac{1}{\pi} \lim_{\varepsilon \rightarrow 0} \Im \log \Delta(E + \ui \varepsilon) + N(0) \, ,
\end{equation}
where $\Theta$ is the Heaviside theta function. Inserting
\eqref{specdetb} leads to
\begin{equation} \label{nb}
N_\mathrm{b}(E) = - \frac{1}{\pi} \Im \log \left[ \frac{\ui}{2} \, \ue^{-\ui k l} \left( 1 - \ue^{2 \ui k l} \right) \right] = \frac{k l}{\pi} - \frac{1}{2}
+ \sum_{n=1}^\infty \frac{1}{\pi n} \sin( 2 n k l) \, ,
\end{equation}
where we have used an expansion of $\log(1-z)$. The result is an exact
trace formula for the spectral staircase. The first two terms in the
final expression are the Weyl terms for the mean staircase
$\overline{N}_\mathrm{b}(E)$, consisting of a leading order volume
term and a boundary correction due to the Dirichlet boundary
conditions. The oscillatory term is a sum over the periodic orbits
(the orbit of length $2l$ and its repetitions)\footnote{Instead of
  adding $N(0)$ in \eqref{staircase_def} (if unknown) one can add a
  constant that is determined by the condition that the constant term
  in the asymptotic expansion of $\overline{N}(E)$ as $E \rightarrow
  \infty$ agrees with the coefficient of $1/E$ in the asymptotic
  expansion of $\Tr G(E)$.}.  A corresponding trace formula for the
density of states can be obtained by differentiating \eqref{nb} with
respect to $E$.

\vspace{1ex}\noindent {\bf Green function for box with source.}  
We follow the same steps for the system with IBC. The Green function
is again obtained from \eqref{green_wronskian}. We choose as left and
right functions linear combinations of the generalised eigenfunctions
\eqref{phik1} $\phi_k$ and $\phi_{-k}$, $k>0$, that satisfy the left
and right boundary conditions, respectively. Using the computer
algebra system Maple we find
\begin{equation} \label{green_box_ibc}
G^{11}(x,y,E) = G_\mathrm{b}(x,y,E) + G_\mathrm{b}(x,0,E) \, \frac{|c|^2}{E - |c|^2 G_\mathrm{b}(0,0,E)} \, G_\mathrm{b}(0,y,E) \, .
\end{equation}
This again agrees with the result for a delta potential after the
replacement $|c|^2/E=\lambda$ \cite{Sie07}. It has also a direct
semiclassical interpretation that can be seen after splitting the zero
length contribution from $G_\mathrm{b}(0,0,E)$ by defining
$G_\mathrm{b}^- = G_\mathrm{b} - G_0$.  One then finds
\begin{equation} \label{green_expand}
\frac{|c|^2}{E - |c|^2 G_\mathrm{b}(0,0,E)} = \frac{{\cal D}}{1 - {\cal D} G_\mathrm{b}^-(0,0,E)} = \sum_{n=0}^\infty ({\cal D} G_\mathrm{b}^-(0,0,E))^n {\cal D} \, ,
\end{equation}
where $\mathcal{D} = 2\ui k b_k$ is the diffraction coefficient,
cf.\ Eq.~\eqref{green_gtd}. After inserting \eqref{green_expand} into
\eqref{green_box_ibc} $G^{11}$ can be interpreted as sum over all regular
trajectories from $y$ to $x$, plus a sum over all diffractive
trajectories from $y$ to $x$ that are diffracted at the source an
arbitrary number of times.

The trace of the resolvent requires also the component in the vacuum
sector. It can be obtained by applying the IBCs \eqref{green_ibcs},
\begin{equation}
G^{00}(E) = \frac{1}{E - |c|^2 G_\mathrm{b}(0,0,E)} \, .
\end{equation}
The trace of the resolvent is evaluated using the resolvent
identity,
\begin{equation}
\int_{-l_1}^{l_2} G_\mathrm{b}(x,z,E) \, G_\mathrm{b}(z,y,E) \, \ud z 
= - \frac{\ud}{\ud E} \, G_\mathrm{b}(z,y,E) \, ,
\end{equation}
leading to
\begin{equation}
\begin{split}
\Tr G(E) & = \int_{-l_1}^{l_2} G^{11}(x,x,E) \, \ud x + G^{00}(E) \\
& = \Tr G_\mathrm{b}(E) - \frac{|c|^2 \frac{\ud}{\ud E} \, G_\mathrm{b}(0,0,E)}{E - |c|^2 G_\mathrm{b}(0,0,E)} + \frac{1}{E - |c|^2 G_\mathrm{b}(0,0,E)} \\
& = \Tr G_\mathrm{b}(E) + \frac{\ud}{\ud E} \log (E - |c|^2 G_\mathrm{b}(0,0,E)) \, . 
\end{split}
\end{equation}
The spectral determinant follows immediately, 
\begin{equation} \label{specdet}
\Delta(E) = \prod_{n=1}^\infty \left( 1 - \frac{E}{E_n} \right) 
= \lim_{\varepsilon \rightarrow 0} \ue^{\int_0^E \Tr G(E + \ui \varepsilon) \, \ud E } 
= \Delta_\mathrm{b}(E) \, \big(E - |c|^2 G_\mathrm{b}(0,0,E)\big) \, \frac{l}{|c|^2 l_1 l_2} \, .
\end{equation}
One can check that the zeros of $\Delta(E)$ coincide with the
solutions of the eigenvalue equations \eqref{boxground} and
\eqref{boxhigher}.  The role of $\Delta_\mathrm{b}$ is to cancel the
poles of $G_\mathrm{b}$ and make the function $\Delta(E)$ entire.

\begin{figure}[t]
\centerline{\includegraphics[width=1.1\textwidth]{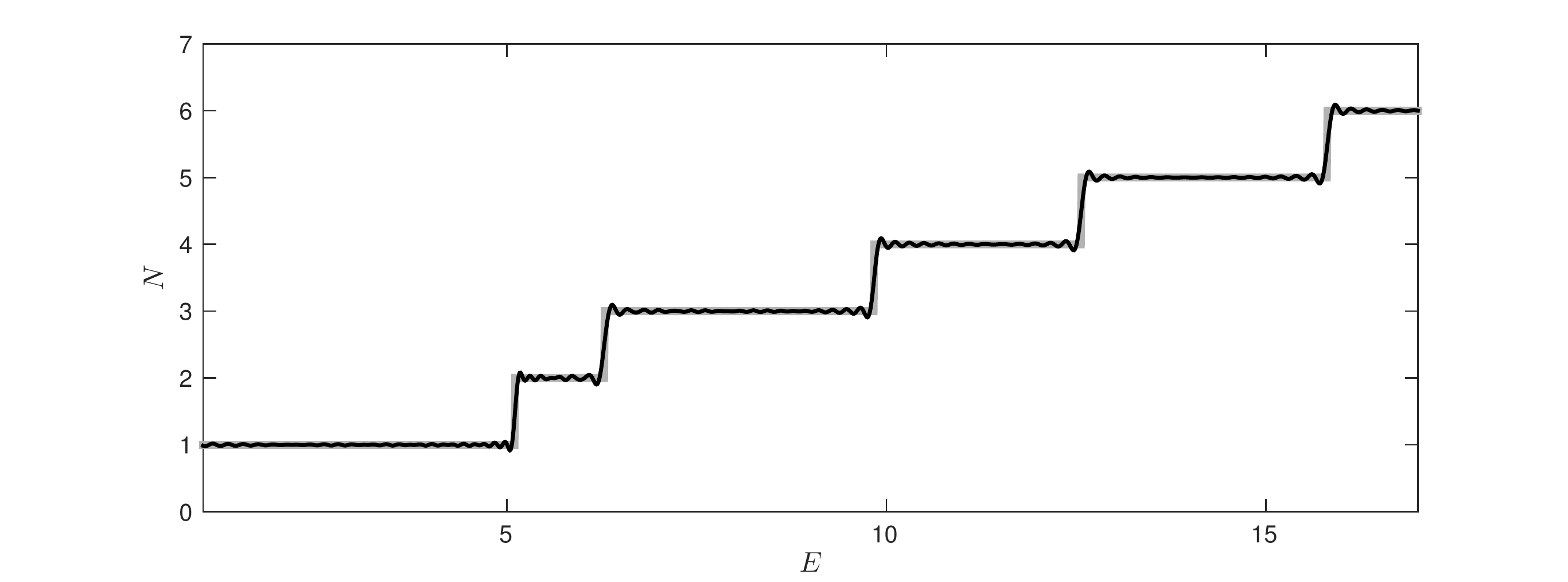}}
\caption{Spectral staircase for a box of length $l=1$ with source in
  the middle, i.e.\ $l_1=l_2=\frac{1}{2}$, and coupling constant
  $c=20$. We compare the exact staircase (grey) to the trace
  formula~\eqref{nofe_final} evaluated using the 2855 shortest
  orbits (black).}
\label{fig:staircase}
\end{figure}

In a last step we calculate the spectral staircase $N(E)$ by applying
\eqref{staircase_def}. With the relation
\begin{equation} \label{factors}
E - |c|^2 G_\mathrm{b}(0,0,E) = ( E - |c|^2 G_0(0,0,E)) \, (1 - {\cal D} G_\mathrm{b}^-(0,0,E)) \, ,
\end{equation}
one obtains
\begin{equation} \label{nofe_gtd}
N(E) =  \frac{k l}{\pi} + \frac{1}{\pi} \arctan \left( \frac{k^3}{\kappa^3} \right) + \sum_{n=1}^\infty \frac{1}{\pi n} \sin( 2 n k l) 
+  \sum_{n=1}^\infty \frac{1}{\pi n} \Im \left( {\cal D} G_\mathrm{b}^-(0,0,E) \right)^n
\end{equation}
This is an exact trace formula for the spectral staircase. The source
contributes a term $1/2+\arctan(k^3/\kappa^3)/\pi$ to the mean
spectral staircase plus a sum over all diffractive orbits of the
system. These are closed orbits that are diffracted at the source an
arbitrary number of times. Their contribution in \eqref{nofe_gtd} is
of the general form that is expected from the geometrical theory of
diffraction \cite{VWR94,PS95,PS95e,BW96,Sie99}. The diffractive orbits
have the role to cancel the steps that are produced by the periodic
orbits at the energy levels of the box without source, and produce
new steps at the energy levels of the box with source.

The contributions of periodic and diffractive orbits can be combined
and simplified by combining the term \eqref{factors} with
$\Delta_\mathrm{b}(E)$ before the expansion into orbits. One then
obtains
\begin{align} \label{nofe_final}
N(E) & = \frac{k l}{\pi} + \frac{1}{\pi} \arctan \left( \frac{k^3}{\kappa^3} \right) + \Im \sum_{n=1}^\infty \frac{(-1)^n}{\pi n} 
\left( \ue^{2 \ui k l + 2 \ui \arctan(k^3/\kappa^3)} + b_k \, \ue^{2 \ui k l_1} + b_k \, \ue^{2 \ui k l_2} \right)^n \notag \\
& = \frac{k l}{\pi} + \frac{1}{\pi} \arctan \left( \frac{k^3}{\kappa^3} \right) + \Im \sum_{n_0,n_1,n_2=0}^\infty\!\!\!\!\!\!\!{}^{'} \frac{(-1)^n}{\pi n} 
\frac{n! \, b_k^{n1+n2}}{n_0! \, n_1! \, n_2!} \ue^{2 \ui k L + 2 \ui n_0 \arctan(k^3/\kappa^3)} \, ,
\end{align}
where on the second line $n=n_0+n_1+n_2$, $L=n_0 l+n_1 l_1+n_2 l_2$,
and the prime indicates the omission of the term with $n_0=n_1=n_2=0$.
Also recall that $b_k$ is defined in \eqref{bk}. A numerical evaluation of
the trace formula for a finite number of orbits is compared to the
exact spectral staircase in Fig.~\ref{fig:staircase}.

\section{IBCs for quantum graphs}
\label{sec:graphs}

Following the seminal work of Kottos and Smilansky
\cite{Kotsmi97,KotSmi99} quantum graphs have become paradigmatic in
the study of quantum chaos and related areas; for overviews see e.g.\
\cite{GnuSmi06,BerKuc13}. In this context one mainly investigates
single-particle quantum mechanics on graphs. The dynamics of a fixed
number of particles interacting by two-particle interactions was
studied in \cite{Har07,Har08} for star graphs and in
\cite{BolKer13a,BolKer13b,BolKer13c} for general graphs. The Fock
space over a quantum graph was introduced for star graphs in
\cite{BelMin06,BelBurMinSor08} and studied for general graphs in
\cite{Sch09}. For some classes of graphs not only boson and fermion
but also anyon statistics are possible
\cite{HarKeaRob10,HarKeaRobSaw14}; the latter we do not discuss
here. In this section we demonstrate that particle creation and
annihilation on Fock space over a graph can be implemented in terms of
IBCs. We discuss three variants of IBCs for quantum graphs along with
their different physical interpretations.

Consider a topological graph $\Gamma$ consisting of $N_v$ vertices,
partially or completely connected by $N_e$ edges. In order to simplify
the notation we restrict the discussion to simple graphs, i.e.\ each
edge connects two different vertices and no two edges connect the same
two vertices. We set $a_{jk}=1$ if the vertices $j$ and $k$ are
connected by an edge, otherwise $a_{jk}=0$. The (symmetric) $N_v
\times N_v$-matrix $A=(a_{jk})$ is the graph's adjacency matrix.
In the following we label each edge by the pair of vertices which it
connects, e.g.\ $(jk)$, $j,k \in 1,\hdots,N_v$, denotes the edge
connecting vertices $j$ and $k$. The graph $\Gamma$ becomes a metric
graph by assigning a length to each edge. We denote the length of edge
$(jk)$ by $l_{(jk)}$, and consequently $l_{(jk)}=l_{(kj)}$.

The one-particle Hilbert space is a union of $L^2$-spaces over the
edges, 
\begin{equation}
\label{eq:L^2-graph}
  \mathcal{H} 
  = \underset{k>j,\,a_{jk}=1}{\bigoplus_{j,k=1}^{N_v}} L^2([0,l_{(jk)}])
\end{equation}
and the wave function in the $n$-particle sector reads
\begin{equation}
  \phi^n_{(j_1 k_1)\cdots(j_n k_n)}(x_1,\hdots,x_n) \, , 
\end{equation}
where $x_\nu \in [0,l_{(j_\nu k_\nu)}]$ is a coordinate on edge
$(j_\nu k_\nu)$ which is zero at the vertex $j_\nu$. It is convenient
to also introduce an alternative coordinate on the same edge, which is
zero at the vertex $k_\nu$. This is easily accommodated within our
notation by
\begin{equation}
  \phi^n_{\cdots(j_\nu k_\nu)\cdots}(\hdots,x_\nu,\hdots)
  = \phi^n_{\cdots(k_\nu j_\nu)\cdots}(\hdots,l_{(j_\nu k_\nu)}-x_\nu,\hdots) 
  \, .
\end{equation}
Wave functions are symmetrised by simultaneously permuting both,
edge labels and coordinates,
\begin{equation}
  (P^S\phi)^n_{(j_1 k_1)\cdots(j_n k_n)}(x_1,\hdots,x_n) 
  = \frac{1}{n!} \sum_{\sigma \in S_n} 
  \phi^n_{(j_{\sigma(1)} k_{\sigma(1)})\cdots(j_{\sigma(n)} k_{\sigma(n)})}
        (x_{\sigma(1)},\hdots,x_{\sigma(n)}) \, . 
\end{equation}
The Laplacian $\Delta^\mathcal{F}$ is the edgewise second derivative
\begin{equation}
  \big(\Delta^\mathcal{F}\phi\big)^n = \sum_{j=1}^n \Delta_j \phi^n \, .
\end{equation}

We now want to allow particle creation and annihilation at exernal
sources. Without loss of generality we only consider sources in
vertices, since sources on edges can easily be accommodated in this
scheme by cutting an edge and adding an additional vertex with valency
two at this position. Hence, as Hamiltonian we choose
\begin{equation}
  H = -\Delta^\mathcal{F} + \sum_{j=1}^{N_v} \overline{c}_j\, a(\delta_j) \, , 
\end{equation}
where in analogy to Eqs.~\eqref{eq:a(f)def} and \eqref{eq:tentative_H}
the annihilator $a(\delta_j)$ annihilates a particle at vertex~$j$,
\begin{equation}
  \big( a(\delta_j) \phi\big)^n_{(j_1k_1)\cdots(j_nk_n)}(x_1,\hdots,x_n)
  = \frac{\sqrt{n+1}}{d_j} \sum_{k\in\Gamma_j} 
    \phi^{n+1}_{(j_1k_1)\cdots(j_nk_n)(jk)}(x_1,\hdots,x_n,0) \, .
\end{equation}
Here the neighbourhood 
\begin{equation}
 \Gamma_j = \{ k \,|\, a_{jk} = 1 \}
\end{equation}
is the set of all vertices which are connected to vertex $j$, and $d_j
= |\Gamma_j|$ is the valency of the vertex.  In each vertex, i.e.\
$\forall\,j=1,\hdots,N_v$, we demand continuity of the wave function,
i.e.\
\begin{equation}
  \phi^n_{(jk)\cdots}(0,\hdots) = \phi^n_{(j\ell)\cdots}(0,\hdots) 
  \quad \forall\, k,\ell \in \Gamma_j \, .
\end{equation}
In analogy to Eq.~\eqref{eq:IBC-model_Fock} the IBCs read
\begin{equation}
  \phi^{n}_{(j_1 k_1)\cdots(j_{n} k_{n})}(x_1,\hdots,x_{n})
  = \frac{\sqrt{n+1}}{c_j} \sum_{k\in\Gamma_j} 
    \left( \partial_{n+1} \phi^{n+1} \right)%
    _{(j_1 k_1)\cdots(j_{n} k_{n})(jk)}(x_1,\hdots,x_{n},0)
  \quad \forall\, j \, .
\end{equation}
We have thus established a quantum field on a graph interacting with
external sources in the vertices.

The minimal model still allowing to describe creation and annihilation
of particles on graphs, i.e.\ the restriction to the 0-1-particle
space $\mathcal{F}^S_1$, reads
\begin{equation}
\label{eq:IBC_graphs-0-1}
\begin{split}
  (H\phi)^1_{jk}(x) &= -{\phi^1_{jk}}''(x) \, , \\
  (H\phi)^0 &= \sum_{j=1}^{N_v} \overline{c}_j\, \phi^1_{jk_j}(0) 
  \quad \forall\, k_j\in\Gamma_j \, , \\
  \text{with IBCs} \quad 
  \phi^0 &= \frac{1}{c_j} \sum_{k\in\Gamma_j} {\phi^1_{jk}}'(0) 
  \quad \forall\, j \, .
\end{split}
\end{equation}
In the limit $c_j\to0$ the Fock space sectors decouple, as expected, and
the IBCs reduce to a Kirchhoff boundary conditions in the one-particle
sector. 

The Hamiltonian \eqref{eq:IBC_graphs-0-1} describes a particle which
moves freely along the edges of a graph, can be annihilated whenever it
reaches a vertex and can eventually be re-created at the same or at a
different vertex. If re-creation of the particle always happened at the
same vertex at which it was previously annihilated, we could
alternatively think of the particle being trapped at this vertex for a
certain period of time.\footnote{We owe this alternative
  interpretation to Stefan Teufel.} Modelling such a situation may be
interesting in its own right, independently of our original motivation
to describe particle creation and annihilation in the context of
quantum field theory. In order to keep track of at which vertex the
particle is trapped we replace the zero-particle wave function $\phi^0
\in \C$ by a vector $\phi^0 \in C^{N_v}$ with components $\phi^0_j$,
one for each vertex. The modified model, which is no longer defined on
$\mathcal{F}^S_1=\C \oplus \mathcal{H}$ but instead on $\C^{N_v}
\oplus \mathcal{H}$, reads
\begin{equation}
\label{eq:IBC_graphs-trapped}
\begin{split}
  (H\phi)^1_{jk}(x) &= -{\phi^1_{jk}}''(x) \, , \\
  (H\phi)^0_j &= \overline{c}_j \, \phi^1_{jk}(0) 
  \quad \forall\, k\in\Gamma_j \, , \\
  \text{with IBCs} \quad 
  \phi^0_j &= \frac{1}{c_j} \sum_{k\in\Gamma_j} {\phi^1_{jk}}'(0) \, .
\end{split}
\end{equation}
This last Hamiltonian is related to so-called quantum decorated
graphs, see e.g.~\cite{Tol08,CheSha13} and references therein. A
decorated graph is a metric graph, with each vertex replaced by a
smooth manifold $\mathcal{M}_j$, $j=1,\hdots,N_v$. The Laplacian on a
decorated graph is then defined on a suitable domain in
\begin{equation}
  \left( \bigoplus_{j=1}^{N_v} L^2(\mathcal{M}_j) \right)
  \oplus \mathcal{H}
\end{equation}
i.e.\ the direct sum of the $L^2$-spaces over the vertex-manifolds and
$\mathcal{H}$, which itself is the direct sum of the $L^2$-spaces over
the edges, see Eq.~\eqref{eq:L^2-graph}. Upon replacing each
$L^2(\mathcal{M}_j)$ by $\C$, we obtain the Hilbert space of
\eqref{eq:IBC_graphs-trapped}.

\section{Conclusions and outlock}

We have shown that the IBC-approach to particle creation and
annihilation developed by Teufel and Tumulka
\cite{TeuTum15a,TeuTum15b} in three spatial dimensions can also be
applied to one-dimensional systems. We have presented an
IBC-Hamiltonian on full Fock space and on a truncated Fock space,
where at most one particle can be created. We have explicitly studied
the characteristics of the 0-1-particle case as the minimal model for
particle creation and annihilation in terms of IBCs. The
IBC-Hamiltonian on full Fock space can also be studied in terms of
explicit solutions; this we will discuss elsewhere. In the
three-dimensional case Teufel and Tumulka introduce a whole family of
different IBC-Hamiltonians. This is also possible in one
dimension. For instance, one can essentially interchange the roles of
$\phi^1$ and ${\phi^1}'$ in \eqref{eq:1D-IBC-Hamiltonian} when
allowing $\phi^1$ to be discontinuous at the origin but demanding
${\phi^1}'(0+)={\phi^1}'(0-)$. More general IBCs, involving linear
combinations of $\phi^1$ and ${\phi^1}'$, also lead to well-defined
models. IBCs in two dimensions can also be defined, which will, e.g.,
allow to study particle creation and annihilation in quantum
billiards. In the long run it will be interesting to see if realistic
quantum field theories can be formulated non-perturbatively using
IBCs. To this end it will be necessary to study quantised gauge fields
and Dirac fermions in terms of IBCs.

\section*{Acknowledgements}
We thank Stefan Teufel and Roderich Tumulka for many discussions and
for sharing their results with us prior to publication. We also
enjoyed helpful discussions with Jonas Lampart and Julian Schmidt.

\begin{appendix}
\section{Three-dimensional IBC-Hamiltonian}
\label{app:3Dibcs}

In Ref.~\cite{TeuTum15a} Teufel and Tumulka present an IBC appropriate
to describe Schrödinger particles in three spatial dimensions which
can be created and annihilated at an external source. They motivate
their choice of IBC by discussing probability conservation on full
Fock space. Here we show that the IBC of Ref.~\cite{TeuTum15a} can
alternatively be motivated along the same lines as its one-dimensional
analogue in Sec.~\ref{sec:ibc-hamiltonian}.

Now the single particle Hilbert space is $\mathcal{H}=L^2(\R^3)$, and
$\mathcal{F}^S = P^S(\bigoplus_{n=0}^\infty \mathcal{H}^{\otimes n})$
as usual. Annihilation and creation operators $a(f)$ and $a^\dag(f)$
are defined as in Eqs.~\eqref{eq:a(f)def} and \eqref{eq:a+(f)def}. The
starting point for our considerations is once more the tentative
Hamiltonian \eqref{eq:tentative_H} where now $y \in \R^3$. Again we
write down the eigenvalue equation $H\phi=E\phi$ in the $n$-particle
sector, cf.~Eg.~\eqref{eq:eigenvalue_n-particle-sector}, integrate in
$x_n$ over the ball $B_{r}(y)=\{ x_n \in \R^3 \, : \, |x_n-y| \leq r
\}$ and subsequently take the limit $r \to 0+$, yielding
\begin{equation}
  -\int_{B_r(y)} (\Delta_n \phi^n)(x_1,\hdots,x_n) \, \ud x_n
  + \frac{c}{\sqrt{n}} \, \phi^{n-1}(x_1,\hdots,x_{n-1}) = 0 \, . 
\end{equation}
By applying Gauß' theorem, the first term can be rewritten as an
integral over the two-sphere,
\begin{equation}
\label{eq:boundary-integral}
  -\int_{B_r(y)} (\Delta_n \phi^n)(x_1,\hdots,x_n) \, \ud x_n 
  = - \int_{S^2} \frac{\partial \tilde{\phi}^n}{\partial r} 
    (x_1,\hdots,x_{n-1},r,\omega) \, r^2 \, \ud\omega \, ,
\end{equation}
where we have introduced spherical coordinates
$(r,\omega)\in[0,\infty) \times S^2$ centered at $y$ through
\begin{equation}
  \tilde{\phi}^n (x_1,\hdots,x_{n-1},r,\omega)
  = \phi^n(x_1,\hdots,x_{n-1},y+r\omega) \, .
\end{equation}
In the limit $r\to0+$ the integral \eqref{eq:boundary-integral}
vanishes for smooth functions $\phi^n$. However, Teufel and Tumulka
show \cite{TeuTum15a} that when constructing the IBC-Hamiltonian one
should include the functions $a \ue^{-\kappa|x-y|}/|x-y|$, $a\in\C$,
$\kappa>0$, in the domain of the one-particle Hamiltonian. For these
we have the non-zero result
\begin{equation}
  -\int_{S^2} \left( \frac{\partial}{\partial r} \, a \, 
    \frac{\ue^{-\kappa r}}{r} \right) r^2 \, \ud\omega
  = 4\pi a (\kappa r + 1) \ue^{-\kappa r} 
  \underset{r\to0+}{\longrightarrow} 4\pi a \, .
\end{equation}
Then again $\lim\limits_{r\to0+} (r \cdot a \ue^{-\kappa r}/r) = a$,
whereas $\lim\limits_{r\to0+} r \phi^n(x_1,\hdots,x_{n+1},y+r\omega) = 0$
for continuous $\phi^n$. Consequently, 
\begin{equation}
  -\lim_{r\to0+} \int_{S^2} \frac{\partial \tilde{\phi}^n}{\partial r} 
    (x_1,\hdots,x_{n-1},r,\omega) \, r^2 \, \ud\omega 
  = 4\pi \lim_{r\to0+} r \phi^n(x_1,\hdots,x_{n-1},y+r\omega) \, , 
\end{equation}
which is the way in which this term is expressed within the IBC of
Ref.~\cite{TeuTum15a}. Now we seem to have another problem. The
Hamiltonian contains the term $a(\delta_y)$ but the functions $a
\ue^{-\kappa|x-y|}/|x-y|$ cannot be evaluated at $x=y$. Here the way
out is to notice that
\begin{equation}
  \phi^n(x_1,\hdots,x_{n-1},y) 
  = \lim_{r\to0+} \frac{\partial}{\partial r} 
    \left( r \, \tilde{\phi}^n(x_1,\hdots,x_{n-1},r,\omega) \right) 
\end{equation}
for smooth functions, and that the right-hand side is still defined
for functions which diverge like $a
\ue^{-\kappa|x_n-y|}/|x_n-y|$. Therefore, we have to redefine the
annihilator according to
\begin{equation}
  \big( a(\delta_y) \phi \big)^n(x_1,\hdots,x_n) 
  = \sqrt{n+1} \lim_{r\to0+} \frac{\partial}{\partial r} 
    \left( r \, \tilde{\phi}^{n+1}(x_1,\hdots,x_n,r,\omega) \right) \, .
\end{equation}
Altogether, the Hamiltonian for the three-dimensional model reads,
cf.~\cite{TeuTum15a}, 
\begin{equation}
\begin{split}
\label{eq:IBC-model_3D}
  &H = -\Delta^\mathcal{F} + \overline{c} \, a(\delta_y) \\
  \text{ with IBC } \ & 
  \phi^n(x_1,\hdots,x_n) = - \frac{4\pi}{c}\sqrt{n+1} 
  \lim_{r\to0+} r \, \phi^{n+1}(x_1,\hdots,x_n,y+r\omega)
  \, .
\end{split}
\end{equation}

\section{Ground state in the limit of vanishing coupling}
\label{app:recover_ground_state}

For vanishing coupling $c=0$ the
IBC-Hamiltonian~\eqref{eq:1D-IBC-Hamiltonian} becomes the free
Hamiltonian $H_\mathrm{free} = -\Delta^\mathcal{F}$, i.e.\
$H_\mathrm{free}(\phi^0,\phi^1)=(0,-{\phi^1}'')$. Its ground state is
the free vacuum $\phi_\mathrm{vac} =
(\phi_\mathrm{vac}^0,\phi_\mathrm{vac}^1) = (1,0)$ with eigenvalue
$E_\mathrm{vac}=0$. The ground state $\phi_\mathrm{g}$ of the
IBC-Hamiltonian for $c\neq0$ is discussed in
Sec.\ref{sec:one_source_on_R}, see Eq.~\eqref{phig}. Naively, one
would expect $\phi_\mathrm{g}$ to approach $\phi_\mathrm{vac}$ in the
limit $c\to0$, except for an arbitrary global phase. However, although
$\lim\limits_{c\to0} E_\mathrm{g} = E_\mathrm{vac}$ and $\|\phi_g\|^2
= 1\ \forall\, c\neq0$ we observe 
\begin{equation}
  \lim_{c\to0} \left( -\tfrac{c}{|c|} \, \phi_\mathrm{g} \right)
  = \Big( \sqrt{\tfrac{2}{3}}, 0 \Big) \, , 
\end{equation}
and in particular $\|\lim\limits_{c\to0} \phi_\mathrm{g}\|^2 = 2/3
\neq \lim\limits_{c\to0} \|\phi_\mathrm{g}\|^2$.

This artefact is a remnant of an infrared divergence which would
appear for the IBC-Hamiltonian on full Fock space. Even for small
couplings the energy of $\phi_\mathrm{vac}$ can be lowered by creating
a small kink in the one-particle sector at the position of the
source. Although $\phi^1_\mathrm{g}$ decays exponentially, the decay
rate decreases with decreasing coupling constant $c$. Physically, this
means that with decreasing coupling the particle cloud surrounding the
source delocalises more and more, and ultimately the particle
partially escapes to infinity. This can be seen as follows. The
expectation value of the particle vanishes, $\langle
\phi^1_\mathrm{g}, x\, \phi^1_\mathrm{g}\rangle_{L^2(\mathbb{R})} = 0\
\forall\, c$, since $\phi^1_\mathrm{g}$ is symmetric. The variance,
however, diverges when the coupling goes to zero, 
\begin{equation}
  \langle \phi^1_\mathrm{g}, x^2\, \phi^1_\mathrm{g}\rangle_{L^2(\mathbb{R})}
  = \frac{\kappa}{3} \int_{-\infty}^\infty \ue^{-2\kappa|x|} \, \ud x 
  = \frac{1}{6\kappa^2} \ 
  \underset{|c|\to0}{\longrightarrow} \ \infty \, .  
\end{equation}

In order to recover the free vacuum in the limit of vanishing
coupling, we have to first add a zero point energy $M>0$ (``rest
mass''), which is only removed after the limit $c\to0$. The modified
IBC-Hamiltonian, cf.\ Eq.~\eqref{eq:1D-IBC-Hamiltonian} then reads 
\begin{equation}
\label{eq:1D-IBC-Hamiltonian_with_mass}
  (H\phi)^1 = -{\phi^1}'' + M \phi^1 \, , \quad 
  (H\phi)^0 = \overline{c} \phi^1(0) \quad \text{with IBC} \quad
  \phi^0 = \frac{1}{c} \Big[{\phi^1}'(x)\Big]_{x=0-}^{x=0+} \, .
\end{equation}
For the ground state we again make the ansatz $\phi^1(x) = A
\exp(-\kappa|x|)$, $\kappa>0$. From the eigenvalue equation in the
one-particle sector we can read off the energy $E=-\kappa^2+M$. In the
zero-particle sector the eigenvalue equation yields
$\overline{c}A=E\phi^0$ $\Leftrightarrow$
$\phi^0=\overline{c}A/(-\kappa^2+M)$ and the IBC reads
$\phi^0=-2A\kappa/c$. The last two conditions can only be satisfied if
$\kappa^2>M$, and then $\kappa$ has to solve
\begin{equation}
\label{eq:condition_kappa_c}
  2\kappa (\kappa^2-M) = |c|^2 \, .
\end{equation}
The normalisation condition $\|\phi\|^2=1$ requires $|A| = \sqrt{\kappa
  \frac{\kappa^2-M}{3\kappa^2-M}}$ implying
\begin{equation}
  |\phi^0|^2 = \frac{2\kappa^2}{3\kappa^2-M} 
  \qquad \text{and} \qquad
  \|\phi^1\|^2 = \frac{\kappa^2-M}{3\kappa^2-M} \, .
\end{equation}
Fulfilling condition~\eqref{eq:condition_kappa_c} under the constraint
$\kappa^2 > M$ requires that $\kappa \to \sqrt{M}+$ when $|c|\to0$,
and thus
\begin{equation}
  |\phi^0|^2 \underset{|c|\to0}{\longrightarrow} 1
  \qquad \text{and} \qquad
  \|\phi^1\|^2 \underset{|c|\to0}{\longrightarrow} 0 \, .
\end{equation}
Hence, except for the undetermined phase, we recover the free vacuum
$(\phi_\mathrm{vac}^0,\phi_\mathrm{vac}^1)=(1,0)$.

\section{Orthonormality and completeness of the eigenstates}
\label{sec:orthocomplete}

{\bf Orthonormality.} 
We show that $\langle \phi_\mathrm{g} , \phi_k \rangle = 0$ and
$\langle \phi_k' , \phi_k \rangle = \delta(k-k')$ for the eigenstates
$\phi_\mathrm{g}$ and $\phi_k$ which are given in (\ref{phig}),
(\ref{phik1}), (\ref{phik0}) and (\ref{bk}).  For the first relation
we consider
\begin{equation}
\overline{\phi_\mathrm{g}^0} \phi_k^0 = - \sqrt{\frac{2}{3\pi}} \; \frac{\sqrt{k^2 \, \kappa^3}}{|k|^3 + \ui \kappa^3} \, ,
\end{equation}
and 
\begin{equation}
\begin{split}
\int_{-\infty}^\infty \overline{\phi_\mathrm{g}^1(x)} \, \phi_{k}^1(x) \, \ud x
&= \int_{-\infty}^\infty \sqrt{\frac{\kappa}{3}} \, \ue^{- \kappa \, |x|} \; \frac{1}{\sqrt{2 \pi}} \left( \ue^{\ui kx} + b_k \ue^{\ui |k||x|} \right) \, \ud x \\
& = \sqrt{\frac{\kappa}{6 \pi}} \left[ \frac{1}{\kappa - \ui k} + \frac{1}{\kappa + \ui k} + \frac{(- 2 \ui \kappa^3)}{(\kappa - \ui |k|) \, (|k|^3 + \ui \kappa^3)}  \right] \\
& = \sqrt{\frac{\kappa}{6 \pi}} \; \frac{2 \kappa \, |k|^3 + 2 \kappa^3 |k|}{(\kappa^2 + k^2) \, (|k|^3 + \ui \kappa^3)} \\
& = \sqrt{\frac{2}{3\pi}} \; \frac{\sqrt{k^2 \, \kappa^3}}{|k|^3 + \ui \kappa^3} \, .
\end{split}
\end{equation}
Adding both results shows that $\langle \phi_\mathrm{g} , \phi_k \rangle = 0$ for all $k \in \mathbb{R}$.

For the inner product between two scattering states we consider
\begin{equation} \label{ss0}
\overline{\phi_{k'}^0} \phi_k^0 
= \frac{2|k k'|}{\pi \, |c|^2} \, \overline{b_{k'}} b_k = \frac{| k' k| \, \kappa^3}{\pi \, (|k|^3 + \ui \kappa^3) \, (|k'|^3 - \ui \kappa^3)} \, ,
\end{equation}
and
\begin{equation} \label{ss1}
\begin{split}
\int_{-\infty}^\infty \overline{\phi_{k'}^1(x)} \, \phi_{k}^1(x) \, \ud x
&= \frac{1}{2 \pi} \int_{-\infty}^\infty  \left[ \ue^{- \ui k'x} + \overline{b_{k' }} \ue^{- \ui |k'||x|} \right] \; \left[ \ue^{\ui kx} + b_k \ue^{\ui |k||x|} \right] \, \ud x \\
& = \delta(k - k') + \frac{1}{2 \pi} \left[ \frac{\ui \, \overline{b_{k' }}}{k - |k'| + \ui \,0+} + \frac{\ui \, \overline{b_{k' }}}{-k - |k'| + \ui \,0+} \right. \\
& \quad \left. + \frac{\ui \, b_k}{-k' + |k| + \ui \,0+} + \frac{\ui \, b_k}{k' + |k| + \ui \,0+} 
+ \frac{2 \ui \, b_k \overline{b_{k' }}}{|k| - |k'| + \ui \,0+} \right] \, ,
\end{split}
\end{equation}
where we have used the identity (see e.g.\ Appendix II of \cite{CDL77-II})
\begin{equation}
  \int_0^\infty \ue^{\ui kx} \, \ud x = \frac{\ui}{k + \ui \,0+} = \pi \delta(k) + {\cal P} \frac{\ui}{k} \, .
\end{equation}
Here ${\cal P}$ stands for principal value. The expression in the square bracket in the final equality of (\ref{ss1}) is even in $k$ and in $k'$.
We assume in the following $k, k' > 0$ and we later use the evenness to extend the result to all real values of $k$ and $k'$.
\begin{equation} \label{ortho1result}
\begin{split}
\int_{-\infty}^\infty \overline{\phi_{k'}^1(x)} \, \phi_{k}^1(x) \, \ud x
& = \delta(k - k') + \frac{\ui}{2 \pi} \; \frac{\left( b_k + \overline{b_{k' }} + 2 b_k \overline{b_{k' }} \right)}{k - k' + \ui \,0+} 
+ \frac{\ui}{2 \pi} \; \frac{\left( b_k - \overline{b_{k' }} \right)}{k + k'} \\
& = \delta(k - k') - \frac{\kappa^3 (k^2 + k k' + {k'}^2)}{2 \pi (k^3 + \ui \kappa^3) ({k'}^3 - \ui \kappa^3)} 
+ \frac{\kappa^3 (k^2 - k k' + {k'}^2)}{2 \pi (k^3 + \ui \kappa^3) ({k'}^3 - \ui \kappa^3)} \\
& =  \delta(k - k') - \frac{k' k \, \kappa^3}{\pi \, (k^3 + \ui \kappa^3) \, ({k'}^3 - \ui \kappa^3)} \, .
\end{split}
\end{equation}
The result is extended to $k, k' \in \mathbb{R}$ by replacing $k$ by $|k|$ and $k'$ by $|k'|$ in the fraction.
Combining the result \eqref{ortho1result} with \eqref{ss0} shows that $\langle \phi_{k'} , \phi_k \rangle = \delta(k - k')$ for all $k, k' \in \mathbb{R}$.

\vspace{1ex}\noindent {\bf Completeness.} 
We showed in section \ref{sec:one_source_on_R} that completeness
requires the three relations
\begin{align}  \label{complete1} 
|\phi_\mathrm{g}^0|^2 + \int_{-\infty}^\infty |\phi_k^0|^2 \, \ud k & = 1 \, , \\ \label{complete2} 
\overline{\phi_\mathrm{g}^0} \, \phi_\mathrm{g}^1(x)  + \int_{-\infty}^\infty \overline{\phi_k^0} \, \phi_k^1(x) \, \ud k & = 0 \, , \\ \label{complete3}
\overline{\phi_\mathrm{g}^1(y)} \, \phi_\mathrm{g}^1(x)  + \int_{-\infty}^\infty \overline{\phi_k^1(y)} \, \phi_k^1(x) \, \ud k  & = \delta(x-y) \, . 
\end{align}
The derivations are similar to that for a delta-potential \cite{Dam75}. Starting with (\ref{complete1}), we obtain
\begin{equation}
|\phi_\mathrm{g}^0|^2 + \int_{-\infty}^\infty |\phi_k^0|^2 \, \ud k
= \frac{2}{3} + \int_{-\infty}^\infty \frac{|k \, b_k|^2}{\pi \, \kappa^3} \, \ud k\\
= \frac{2}{3} + \frac{1}{\pi} \int_{-\infty}^\infty \frac{k^2 \, \kappa^3}{k^6 + \kappa^6} \, \ud k\\ 
= 1 \, .
\end{equation}
For the second relation (\ref{complete2}) we consider
\begin{equation} \label{comp2a}
\overline{\phi_\mathrm{g}^0} \, \phi_\mathrm{g}^1(x)  = - \frac{\sqrt{2 \kappa}}{3} \, \ue^{- \kappa \, |x|} \, \ue^{\ui \varphi_c} \, ,
\end{equation}
and
\begin{equation}
\int_{-\infty}^\infty \overline{\phi_k^0} \, \phi_k^1(x) \, \ud k = \int_{-\infty}^\infty \sqrt{\frac{k^2 \kappa^3}{2 \pi^2}} \, \frac{\ue^{\ui \varphi_c}}{|k|^3 - \ui \kappa^3} \, 
\left[ \ue^{\ui \, k \, x} + \frac{- \ui \kappa^3}{|k|^3 + \ui \kappa^3} \, \ue^{\ui \, |k| \,|x|} \right] \, \ud k \, .
\end{equation}
Applying Euler's formula $\ue^{\ui k x} = \cos(k x) + \ui \sin(k x)$
shows that only the cosine term contributes and that the expression is
even in $x$.  Hence we may assume $x \geq 0$ in the following and
later replace $x$ by $|x|$ to extend the result to all $x \in
\mathbb{R}$. Furthermore, we use evenness in $k$ to integrate only
over positive values of $k$. After inserting Euler's formula also for
the second exponential in the square bracket we obtain
\begin{equation}
\begin{split}
\int_{-\infty}^\infty \overline{\phi_k^0} \, \phi_k^1(x) \, \ud k
& = \frac{\sqrt{2 \kappa^3}}{\pi} \, \ue^{\ui \varphi_c} \, \int_0^\infty \frac{k}{k^6 + \kappa^6} \, \left[ k^3 \cos(k x) + \kappa^3 \sin(k x) \right] \, \ud k \\
& = \frac{\sqrt{2 \kappa^3}}{4 \pi} \, \ue^{\ui \varphi_c} \int_{-\infty}^\infty \frac{k}{k^6 + \kappa^6} \, 
\left[ (k^3 - \ui \kappa^3) \ue^{\ui k x} + (k^3 + \ui \kappa^3) \ue^{-\ui k x} \right] \, \ud k \\
& = \frac{\sqrt{2 \kappa^3}}{2 \pi} \, \ue^{\ui \varphi_c} \int_{-\infty}^\infty \frac{k}{k^3 + \ui \kappa^3} \, \ue^{\ui k x} \, \ud k \\
& = \frac{\sqrt{2 \kappa}}{3} \, \ue^{- \kappa \, x} \, \ue^{\ui \varphi_c} \, .
\end{split}
\end{equation}
The integral was evaluated by contour integration around the pole in the upper half plane at $k= \ui \kappa$. The result is extended 
to $x \in \mathbb{R}$ by replacing $x$ by $|x|$. Combining the result with (\ref{comp2a}) confirms the second relation (\ref{complete2}).

For the third relation (\ref{complete3}) we consider
\begin{equation} \label{comp3a}
\overline{\phi_\mathrm{g}^1(y)} \, \phi_\mathrm{g}^1(x) = \frac{\kappa}{3} \, \ue^{- \kappa (|x| + |y|)} \, ,
\end{equation}
and
\begin{equation} \label{comp3b}
\begin{split}
\int_{-\infty}^\infty \overline{\phi_k^1(y)} \, \phi_k^1(x) \, \ud k  & = \frac{1}{2 \pi} \int_{-\infty}^\infty  \left[ \ue^{\ui kx} + b_k \ue^{\ui |k||x|} \right] \,
 \left[ \ue^{- \ui k y} + \overline{b_k} \ue^{-\ui |k| |y|} \right] \, \ud k \\
 & = \delta(x - y) + \frac{\kappa^3}{2 \pi} \left( I_1(x,y) + I_2(x,y) + I_3(x,y) \right) \, ,
\end{split}
\end{equation}
where
\begin{equation} \label{threeis}
I_1(x,y) = \ui \int_{-\infty}^\infty \frac{1}{|k|^3 - \ui \kappa^3} \, \ue^{\ui k x} \ue^{- \ui |k| |y|} \, \ud k \, , \quad
I_3(x,y) = \int_{-\infty}^\infty \frac{\kappa^3}{k^6 + \kappa^6} \, \ue^{\ui |k| |x|} \ue^{- \ui |k| |y|} \, \ud k \, ,
\end{equation}
and $I_2(x,y) = \overline{I_1(y,x)}$. Using Euler's formula $\ue^{\ui
  k x} = \cos(k x) + \ui \sin(k x)$ for $I_1(x,y)$ shows that only the
cosine term contributes. It shows further that $I_1(x,y)$ is even in
$x$ and in $y$ and that its integrand is even in $k$. The same holds
for $I_2(x,y)$ and $I_3(x,y)$.  We assume $x,y \geq 0$ in the
following and restrict the integrals to positive values of $k$. We
bring all fractions onto the denominator $(k^6 + \kappa^6)$ and obtain
\begin{align}
I_1(x,y) & = \int_0^\infty \frac{2 \cos(k x)}{k^6 + \kappa^6} \left[ \left( k^3 \sin(ky) - \kappa^3 \cos(ky) \right) + \ui \, \left( k^3 \cos(ky) + \kappa^3 \sin(ky) \right) \right] \, \ud k \, , \notag \\
I_2(x,y) & = \int_0^\infty \frac{2 \cos(k y)}{k^6 + \kappa^6} \left[ \left( k^3 \sin(kx) - \kappa^3 \cos(kx) \right) - \ui \, \left( k^3 \cos(kx) + \kappa^3 \sin(kx) \right) \right] \, \ud k \, , \notag \\
I_3(x,y) & = \int_0^\infty \frac{2 \kappa^3}{k^6 + \kappa^6} \left[ \cos(k(x-y)) + \ui \, \sin(k(x-y)) \right] \, \ud k \, .
\end{align} 
After adding all three terms the imaginary part vanishes and the real part gives
\begin{align} \label{threeisresult}
I_1(x,y) + I_2(x,y) + I_3(x,y) & = \int_0^\infty \frac{2}{k^6 + \kappa^6} \left[ k^3 \sin(k(x+y)) - \kappa^3 \cos(k(x+y)) \right] \, \ud k \notag \\
& = - \frac{\ui}{2} \int_{-\infty}^\infty \left[ \frac{k^3 - \ui \kappa^3}{k^6 + \kappa^6}  \ue^{\ui k (x+y)}
                                                       - \frac{k^3 + \ui \kappa^3}{k^6 + \kappa^6}  \ue^{-\ui k (x+y)} \right] \, \ud k \notag \\
& = -\ui \int_{-\infty}^\infty \frac{1}{k^3 + \ui \kappa^3} \, \ue^{\ui k (x+y)} \, \ud k \\
& = - \frac{2 \pi}{3 \kappa^2} \, \ue^{- \kappa (x+y)} \, . \notag
\end{align}
The integral was evaluated by contour integration in the upper half plane around the pole at $k = \ui \kappa$. The result can be extended to $x,y \in \mathbb{R}$
by replacing $x$ by $|x|$ and $y$ by $|y|$. Combining the result with (\ref{comp3a}) and (\ref{comp3b}) shows (\ref{complete3}).

\section{Time evolution kernel}
\label{sec:propagator}

This section contains a derivation of the time evolution kernel of the
IBC-Hamiltonian~(\ref{eq:1D-IBC-Hamiltonian}), based on an
eigenfunction expansion. Alternatively, the kernel can also be
obtained from a Fourier-Laplace transform of the Green function in
(\ref{G11}) and (\ref{G01G10G00}), leading to the same result.

The eigenfunction expansion of the time evolution operator
  $K(T)=\ue^{-\ui HT}$ has the form
\begin{equation*}
  K(T) = \phi_\mathrm{g} \langle\phi_\mathrm{g},\cdot\rangle \, \ue^{\ui \kappa^2 T} 
  + \int_{-\infty}^\infty \phi_k \langle\phi_k,\cdot\rangle \, \ue^{-\ui k^2 T} \ud k \, , 
\end{equation*}
and the eigenstates are given in (\ref{phig}), (\ref{phik1}) and
(\ref{phik0}).

We consider first the 11-sector. After inserting the eigenfunctions
$\phi_\mathrm{g}^1$ and $\phi_k^1$ we obtain the integral kernel 
\begin{equation} \label{k11start}
K^{11}(x,y,T) = \frac{\kappa}{3} \ue^{-\kappa(|x|+|y|)} \ue^{\ui \kappa^2 T} + K_0^{11}(x,y,T) + \frac{\kappa^3}{2 \pi} 
(I_1(x,y,T) + I_2(x,y,T) + I_3(x,y,T)) ,\end{equation}
where
\begin{equation}\label{eq:single_particle_free_K}
K_0(x,y,T) = \frac{1}{2 \pi} \int_{-\infty}^\infty \ue^{- \ui k^2 T} \ue^{\ui k (x-y)} \, \ud k = \frac{1}{\sqrt{4 \pi \ui T}} \exp\left( - \frac{(x - y)^2}{4 \ui T} \right) \, ,
\end{equation}
is the free time evolution kernel in single-particle quantum
mechanics, and
\begin{equation}
\begin{split}
I_1(x,y,T) & =  \ui \int_{-\infty}^\infty \frac{1}{|k|^3 - \ui \kappa^3} \, \ue^{\ui k x} \ue^{- \ui |k| |y|} \ue^{- \ui k^2 T} \, \ud k \, , \\
I_2(x,y,T) & = -\ui \int_{-\infty}^\infty \frac{1}{|k|^3 + \ui \kappa^3} \, \ue^{-\ui k y} \ue^{\ui |k| |x|} \ue^{- \ui k^2 T}\, \ud k \, , \\
I_3(x,y,T) & = \int_{-\infty}^\infty \frac{\kappa^3}{k^6 + \kappa^6} \, \ue^{\ui |k| |x|} \ue^{- \ui |k| |y|} \ue^{- \ui k^2 T}\, \ud k \, .
\end{split}
\end{equation}
Note that the functions $I_j$ differ from those in (\ref{threeis})
only by the additional term $\ue^{- \ui k^2 T}$ in the integrand.  The
calculations are completely analogous to those from (\ref{threeis}) to
(\ref{threeisresult}) and lead to
\begin{equation} \label{iresults}
I_1(x,y,T) + I_2(x,y,T) + I_3(x,y,T) = -\ui \int_{-\infty}^\infty \frac{1}{k^3 + \ui \kappa^3} \, \ue^{\ui k (|x|+|y|)} \ue^{-\ui k^2 T} \, \ud k \, .
\end{equation}
The integral can be evaluated after applying the partial fraction expansion
\begin{equation}
\frac{1}{k^3 +\ui \kappa^3} = - \frac{1}{3 \kappa^3} \left[ \frac{\kappa_0}{k - \ui \kappa_0} + \frac{\kappa_1}{k - \ui \kappa_1} + \frac{\kappa_{2}}{k - \ui \kappa_{2}} \right] \, , 
\end{equation}
where $\kappa_j = \kappa \, \exp(2 \pi \ui j/3)$ for $j=0,1,2$. Note that a useful formula is (for $n \in \mathbb{Z}$)
\begin{equation} \label{kapparel}
\sum_{j=0}^2 \kappa_j^n = \begin{cases} 3 \kappa^n & \text{if $n$ is divisible by 3,} \\ 0 & \text{otherwise.} \end{cases} 
\end{equation}
After inserting the partial fraction expansion into (\ref{iresults}), the resulting integral can be evaluated with the formula
\begin{equation} \label{integral}
\int_{-\infty}^\infty \frac{\exp(-\beta z^2 + \ui \alpha z)}{z \mp \ui \gamma}  \, \ud z = \pm \ui \pi \, \exp(\beta \gamma^2 \mp \alpha \gamma) \;
\operatorname{erfc} \left( \sqrt{\beta} \gamma \mp \frac{\alpha}{ 2 \sqrt{\beta}} \right) \, ,
\end{equation}
where $\Re \alpha > 0$, $\Re \beta \ge 0$, $\Re \gamma > 0$ and erfc is the complementary error function. 
Equation (\ref{integral}) can be obtained, for example, by using
$$ \frac{1}{z \mp \ui \gamma} = \pm \ui \int_0^\infty \ud t \, \exp(\mp \ui (z \mp \ui \gamma)t) \, .$$
After applying the integral (\ref{integral}) one finally obtains the result for the time evolution kernel
\begin{equation} \label{k11final}
K^{11}(x,y,T) = K^{11}_0(x,y,T) + \sum_{j=0}^2 \frac{\kappa_j}{6} \, \ue^{\ui \kappa_j^2 T} \, \ue^{ - \kappa_j (|x|+|y|)} \, 
\operatorname{erfc} \left(\frac{|x|+|y|}{2 \sqrt{\ui T}} - \kappa_j \sqrt{\ui T} \right) \, .
\end{equation}
The ground state contribution in (\ref{k11start}) has been included here in the $j=0$ term. Equation (\ref{k11final})
is the final result for the time evolution kernel. It can be expressed in an alternative form that can be obtained after inserting
the integral representation of the error-function and changing the integration variable
\begin{equation}
K^{11}(x,y,T) = K_0(x,y,T) + \sum_{j=0}^2 \frac{\kappa_j}{3} \int_0^\infty \ue^{\kappa_j u} \, K_0(|x|+|y|+u,0,T) \, \ud u \, .
\end{equation}
There are again similarities to the time evolution kernel for a delta-potential \cite{GS86,Bli88}.

The expressions for the kernel in the other sectors can be obtained
from the eigenfunction expansion in these sectors, or alternatively by
applying the IBCs to $K^{11}(x,y,T)$. We give here only the results
\begin{equation}
\begin{split}
\label{k10k01k00}
K^{10}(x,T) & = - \sum_{j=0}^2 \frac{\kappa_j^2}{3 \overline{c}} \, \ue^{\ui \kappa_j^2 T} \, \ue^{ - \kappa_j |x|} \, 
\operatorname{erfc} \left(\frac{|x|}{2 \sqrt{\ui T}} - \kappa_j \sqrt{\ui T} \right) \, , \\
K^{01}(y,T) & = - \sum_{j=0}^2 \frac{\kappa_j^2}{3 c} \, \ue^{\ui \kappa_j^2 T} \, \ue^{ - \kappa_j |y|} \, 
\operatorname{erfc} \left(\frac{|y|}{2 \sqrt{\ui T}} - \kappa_j \sqrt{\ui T} \right) \, , \\
K^{00}(T) & 
= \frac{1}{3} \sum_{j=0}^2 \ue^{\ui \kappa_j^2 T} \, 
\operatorname{erfc} (- \kappa_j \sqrt{\ui T} ) \, .
\end{split}
\end{equation}

In Sec.~\ref{sec:one_source_on_R} we use the long-time behaviour of
$K^{00}(T)$. It can be obtained from the asymptotics of the
erfc-function
\begin{equation} \label{erfcasym}
\operatorname{erfc}(z) = \frac{\exp(-z^2)}{\sqrt{\pi} z} \left( 1 - \frac{1}{(2 z^2)} + \frac{1\cdot 3}{(2 z^2)^2} + \ldots \right) \quad \text{as} \quad 
|z| \rightarrow \infty , \, |\arg z| < \frac{3 \pi}{4} \, .
\end{equation}
In the remaining sector of $\arg z$ one has to add a $2$ to the
asymptotic expansion. Using (\ref{erfcasym}) and (\ref{kapparel}) one
finds
\begin{equation} \label{k00asympt}
  K^{00}(T) = \frac{2}{3} \, \ue^{\ui \kappa^2 T} + \mathcal{O}(T^{-3/2}) 
  \quad \text{as} \quad T \rightarrow \infty \, .
\end{equation}

The short-time behaviour of the time evolution kernel is supposed to
reveal the underlying classical dynamics, see e.g.\ \cite{Gut90}. For
$K^{00}$ a Taylor expansion yields
$K^{00}(T)=1+\mathcal{O}(T^{3/2})$. The phases of the arguments of the
error functions in the other three components all approach
$-\frac{\pi}{4}$ for small $T$. Thus, we can use the asymptotics
\eqref{erfcasym}, and performing the $j$-sum with the help of
Eq.~\eqref{kapparel} yields
\begin{equation}
\begin{split}
  K^{11}(x,y,T) &= K_0(x,y,T) - \frac{4\ui|c|^2\,T^3}{(|x|+|y|)^3}\,
  \frac{\exp\left(-\frac{(|x|+|y|)^2}{4\ui T}\right)}{\sqrt{4\pi\ui T}}\,
  (1+\mathcal{O}(T)) \, ,\\ 
  K^{10}(x,T) &= \frac{2c\,T^2}{|x|^2}\,
  \frac{\exp\left(-\frac{|x|^2}{4\ui T}\right)}{\sqrt{4\pi\ui T}}\,
  (1+\mathcal{O}(T)) \, ,\\ 
  K^{01}(y,T) &= \frac{2\overline{c}\,T^2}{|y|^2}\,
  \frac{\exp\left(-\frac{|y|^2}{4\ui T}\right)}{\sqrt{4\pi\ui T}}\,
  (1+\mathcal{O}(T)) \, .
\end{split}
\end{equation}
These results can be understood in a similar way as the Green function
in Sec~\ref{sec:one_source_on_R}.  $K^{11}$, in addition to the direct
term $K_0$, contains a diffractive contribution which can be
associated with a path of length $|x|{+}|y|$. We interpret this term
as coming from a particle moving from $y$ to the origin, where it is
annihilated, subsequently re-created, and which then moves on to
$x$. The dependence on the coupling, $|c|^2$ is consistent with
annihilation, proportional to $\overline{c}$, and subsequent
re-creation, yielding a factor $c$. We also observe that the
diffractive contribution decreases with increasing $(|x|{+}|y|)/T$,
the mean velocity along the path. Thus, only slow particles couple
strongly to the source. Similar interpretations apply for $K^{10}$ and
$K^{01}$.

\end{appendix}

\bibliographystyle{my_unsrt}
\bibliography{literatur_ibc}

\end{document}